%% file: main.tex
\pdfoutput=1

 \PassOptionsToPackage{table}{xcolor}
\documentclass[10pt,journal,compsoc]{IEEEtran}
\ifCLASSOPTIONcompsoc
  \usepackage[nocompress]{cite}
\else
  \usepackage{cite}
\fi

\usepackage[dvipsnames]{xcolor}
 \usepackage{dblfloatfix}
 
\usepackage{cite}
\usepackage{multirow}
\usepackage{amsmath}
\usepackage{amsfonts}
\usepackage{balance}

\usepackage{enumitem}
\usepackage{bigstrut}
\usepackage{wrapfig}
\usepackage{color,soul}
\usepackage{url}
\usepackage{fourier}
\usepackage{booktabs}
\usepackage[flushleft]{threeparttable}
\usepackage[usestackEOL]{stackengine}
\usepackage{graphicx}
\usepackage[makeroom]{cancel}
\usepackage{tikz}
\usetikzlibrary{angles, positioning}

\usepackage[framemethod=TikZ]{mdframed}

\newcommand{\BLUE}{\color{black}}
\newcommand{\BLACK}{\color{black}}
\newcommand*{\priority}[1]{\begin{tikzpicture}[scale=0.12]%
    \draw (0,0) circle (1);
    \fill[fill=black] (0,0) -- (90:1) arc (90:90-#1*3.6:1) -- cycle;
    \end{tikzpicture}}
    
\newcommand{\yesCircle}{\priority{100}}
\newcommand{\noCircle}{\priority{0}}
\newcommand{\halfCircle}{\priority{50}}

\newcommand{\bi}{\begin{itemize}}%[leftmargin=0.4cm]}
\newcommand{\ei}{\end{itemize}}
\newcommand{\be}{\begin{enumerate}}
\newcommand{\ee}{\end{enumerate}}
\newcommand{\tion}[1]{\S\ref{sect:#1}}
\newcommand{\fig}[1]{Fig.~\ref{fig:#1}}
\newcommand{\tab}[1]{Table~\ref{tab:#1}}

\newcommand{\eq}[1]{Eq.~\ref{eq:#1}}

\newenvironment{result}[2]
{\vspace{0.8em}\begin{mdframed}[backgroundcolor=gray!10,roundcorner=0pt] \textbf{\textit{\underline{Conclusion \##1: }}} #2}{\end{mdframed} \vspace{0.8em}}

\newenvironment{answer}[2]
{\vspace{0.8em}\begin{mdframed}[backgroundcolor=gray!10,roundcorner=0pt] \textbf{\textit{\underline{Answer \##1: }}} #2}{\end{mdframed} \vspace{0.8em}}
    
\definecolor{light-gray}{gray}{0.90}

\newcolumntype{C}[1]{>{\centering\let\newline\\\arraybackslash\hspace{0pt}}m{#1}}
\newcolumntype{H}{>{\setbox0=\hbox\bgroup}c<{\egroup}@{}}

\begin{document}
\title{Building Very Small Test Suites (using {\sc Snap})}

\author{Jianfeng~Chen,
~Xipeng~Shen,~\IEEEmembership{Senior Member,~IEEE},
        and~Tim~Menzies,~\IEEEmembership{Fellow,~IEEE}
\IEEEcompsocitemizethanks{\IEEEcompsocthanksitem
E-mail: jchen37@ncsu.edu, xshen5@ncsu.edu, timm@ieee.org 
\IEEEcompsocthanksitem J. Chen is  at Facebook. X. Shen and T. Menzies are   at NC State University.}
\thanks{Manuscript received March 23, 2020; revised XXX.}}

%\markboth{Journal of \LaTeX\ Class Files,~Vol.~14, No.~8, August~2015}%
%{Shell \MakeLowercase{\textit{{\it et al.}}}: Bare Demo of IEEEtran.cls for Computer Society Journals}

\IEEEtitleabstractindextext{%
\begin{abstract}
\BLUE
Software is now so large and complex that additional  architecture  is  needed  to  guide  
  theorem provers as they try to generate   test suites. 
For example,  the
 {\sc Snap}   test suite generator (introduced in this paper)  
 combines the Z3 theorem prover with the following   tactic:
sample   around   the   average   values   seen   in   a   few randomly selected valid tests.
This tactic is remarkably effective.
For  27 real-world programs  with up to half a million variables,
  {\sc Snap} 
 found test suites  which were   10 to 750 smaller times than those found by the
 prior state-of-the-art.
 Also,  
  {\sc Snap}  ran orders of magnitude faster
 and (unlike prior work)   generated 100\% valid tests.
 \BLACK
\end{abstract}
}
\maketitle
\IEEEdisplaynontitleabstractindextext
\IEEEpeerreviewmaketitle
\IEEEraisesectionheading{\section{Introduction}\label{sect:intro}}

%Snap was developed after experimenting with the {\em QuickSampler} method proposed 
%by
%Dutra {\it et al.}  at ICSE'18~\cite{dutra2018efficient}.

\BLUE
When changing software, it is useful
to test if the new work damages
old functionality. For this reason,
testing and re-testing code 
is widely applied in both open-source projects and closed-source projects~\cite{fazlalizadeh2009prioritizing, lu2009introReg, mahajan2015intorReg}.
% When developers extend   a code base, such tests let them  check that their new work does not harm old functionality.
% Such  tests means that developers can  detect more faults within a period of time and 
%     start to fix software bugs earlier than usual.
% Hence, better tests enables faster code modification~\cite{fazlalizadeh2009prioritizing, lu2009introReg, haidry2013TCP}. 
 But generating test suites can be difficult since good test suite generators must struggle to achieve    five goals:
\be
\item
 Terminate quickly;
\item Scale to  large programs;
\item Return small   test suites
that  contain valid tests;
\item Cover most   program branches;
\item
Avoid excessive
redundancy  in the tests.
\ee
For example,  the
{\em QuickSampler}  system ~\cite{dutra2018efficient} presented at ICSE'18 runs faster than prior work~\cite{ermon2012uniform,chakraborty2015parallel} for larger programs and   finds test suites with more  valid tests~\cite{dutra2018efficient}. {\em QuickSampler} uses the  heuristic that
 ``valid tests can   be built by combining    other  valid tests';
 e.g.  a new test can be built
 from valid tests $a,b,c$    using  $\oplus$   (``exclusive or''):
\begin{equation}\label{eq:one}
d=c\oplus(a\oplus b)
\end{equation}
This heuristic is useful since    ``exclusive or'' is    faster  than, say, running a theorem prover. 
But   
 heuristics are, by definition,   just guesses (albeit, good ones). Therefore it is hardly surprising
 that   heuristics like \eq{one} can introduce new problems. For example:
  \bi
  \item {\em Redundancy:}  In one sample of 10 million
tests generated from the   {\it blasted\_case47}, a benchmark in this paper.
{\em QuickSampler} only  found 26,000 unique valid solutions.   
That is, 99\% of the tests
were repeating other tests.  
\item {\em Credibility:} A  test suite becomes {\em more}
credible as the number of valid tests  {\em increases}. 
As shown below, 30\% (on average) of tests found by  {\em QuickSampler}  are 
not credible.
\item {\em Minimality:} 
Our  {\sc Snap} tool
 finds test  which are   10 to 750 smaller than those found by  {\em QuickSampler}.
  \ei
  \BLACK
 % Dutra {\it et al.} used this heuristic in their {\em QuickSampler} tool
% to generate 
% valid tests in less than  that 10 $\mu s$ (in 
% the majority of benchmarks). Hence, at least for the problems
% they explored at ICSE'18, {\em QuickSampler} finds millions of tests within two hours.
 This paper introduces {\sc Snap}. We show that {\sc Snap}  runs  orders of magnitude faster
  than  {\em QuickSampler} and its tests are  more credible (since 100\% of its tests are valid).
  
 {\sc Snap} was designed as follows.
{\em QuickSampler} generates   tests by converting   code into a logical formula (discussed later in this paper). The presence of many repeated tests is hence an indication that there is much repeated structure in the formula (as well
as in  the code). 
The starting point for this paper was the conjecture that, 
when repeated structures exist,  the common settings seen in a small $M$ sample may also be the common settings in a large $N\gg M$ sample. If so, the space of tests can be explored
using the following  ``{\sc Snap} tactic'':
 
\begin{quote}
{\em Sample around the average values seen in  a few randomly selected valid tests.}
\end{quote}
 To evaluate this tactic,
 this paper  combines
 \eq{one} with 
the Z3 theorem prover with
the {\sc Snap} tactic.  
The results of that combined tool were then compared to
the more extensive search offered by {\em QuickSampler}.
In that comparison, we asked:

{\bf RQ1: How much faster is the {\sc Snap} tactic? }   
\begin{answer}{1}
~{\sc Snap} terminated  10 to 3000 times faster than {\em QuickSampler} (median to max).
\end{answer}

{\bf RQ2: Does the    {\sc Snap} tactic find fewer
test cases? } 
\begin{answer}{2}
~Test cases from  {\sc Snap} were    10 to 750 times smaller than  those from
  {\em QuickSampler} (median to max).
\end{answer}
\begin{figure*}[!b]

\begin{mdframed}[backgroundcolor=blue!05, linecolor=black, roundcorner=10pt]
\begin{minipage}{2.5in}
\begin{verbatim}
1 int mid(int x, int y, int z) {
2  if (x < y) {
3    if (y < z) return y;
4    else if (x < z) return z;
5    else return x;
6  } else if (x < z) return x;
7  else if (y < z) return z;
8  else return y; }
\end{verbatim}

The code above   has the six branches shown below.
Each branch is a logical constraint  \mbox{$C_1 \vee C_2 \vee C_3\ldots\vee C_6$}.
 A valid test  selects
    {\tt x,}

\end{minipage}
\hspace{0.5in}
\begin{minipage}{3.5in}
{\tt y, z} such that it satisfies
these constraints.
\begin{verbatim}
path 1: [C1: x < y < z] L2->L3  
path 2: [C2: x < z < y] L2->L3->L4 
path 3: [C3: z < x < y] L2->L3->L4->L5 
path 4: [C4: y < x < z] L2->L6 
path 5: [C5: y < z < x] L2->L6->L7 
path 6: [C6: z < y < x] L2->L6->L7->L8
\end{verbatim}

%Note: 1) as \texttt{x}, \texttt{y} and \texttt{z} are integers, we could use (say) 7 bits to representing them;
%2) the arithmetic conditions, ie. the SMT form(Satisfiability Modulo Theories)~\cite{barrett2018satisfiability} can be translated into norm formulas
via SMT conversion tools~\cite{finke2015equisatisfiable}. By convention, the disjunction $\vee C_i$ is
transformed into the conjunction normal form (CNF)  $C_1' \wedge C_2' \ldots$.
% \item Final output of the automated theorem proving is the conjunctive  normal form (CNF). 
A valid assignment to the CNF, i.e. the assignment that fulfills all clauses, is corresponding to a test case, covering some branch of code.
%  When translated into the input required for Z3,
%  these conjunctions look like:
% \begin{verbatim}
% 1 p cnf 11511 41411
% 2 ...
% 3 -11507 11510 0
% 4 -11510 11504 11507 11502 0
% 5 ...
% \end{verbatim}

% Line 1 indicates that that this CNF expression has 11511 variables which are shared
% between  41411 clauses.
% Remaining lines the  41411 clauses (and a zero at end of line denotes end-of-clause).
% For example,
% line 2-5 can  be read as
% $$\ldots\pmb{\wedge}(\neg 11504\vee 11510)\pmb{\wedge}(\neg 11510\vee 11504\vee 11507\vee 11502)\pmb{\wedge}\ldots$$
\end{minipage}
\end{mdframed}
\caption{A script of C programming can be translated into CNF
(conjunctive normal form).}\label{fig:codedemo}
\end{figure*}

{\bf RQ3: How ``good'' are the tests found via the {\sc Snap} tactic? } 
For ``credibility'' defined as the 
percent of valid tests;
and ``diversity'' defined as percent  
code branches covered:
\begin{answer}{3}
        Usually,    {\sc Snap}'s tests are much more credible and 
very nearly as diverse
as those from {\em QuickSampler}.
\end{answer}

% {\color{red} % XXX TOFIX
% The rest of this paper is structured as follows:
% \tion{background} introduces some related works in solving this problem.
% \tion{aboutSnap} shows the core algorithm of {\sc Snap}. 
% \tion{exp} addresses the details of experiments and the study cases.
% \tion{results} discusses the experiment results.
% Following that, \tion{further} and \tion{conclusion} have further discussion and conclusions.
% }
The rest of this paper is structured as follows.
The next section  discusses why we want to shrink test suites
and how to do that using   SAT solvers. Then, after describing  {\scshape {\sc Snap}},
we run experiments on the same 
case studies used in the   ICSE'18 {\em QuickSampler} paper. Finally, we discuss threats to validity.

Note that an open-source version of  
{\sc Snap} (and all SE models used in this paper) is freely available
on-line\footnote{  \url{http://github.com/ai-se/SatSpaceExpo}}.

\newpage
\section{Background}
\BLUE
\subsection{Why Reduce Test Suite Size?}\label{sect:why}

This paper evaluates the {\sc {\sc Snap}} test suite generator using
the five goals described above:
i.e. {\em runtime},   {\em scalability},   {\em redundancy}, {\em credibility} and {\em minimality}.
But before that,  we offer
  general notes   on the important of test suite minimization.

% The {\sc Snap} tool described below generated test suites that are orders of magnitude smaller than those seen in prior results.
% This is useful, for many reasons.

\begin{table*}[!b]
\label{tab:refs}
\caption{{\sc Snap} and its related work for solving theorem proving constraints via sampling.}\label{tbl:refs}
\begin{center}
\begin{tabular}{ccp{0.7in}cC{1in}C{0.7in}C{1in}}
\toprule
\textbf{Reference} & \textbf{Year} & \textbf{Citation} & \textbf{Sampling methodology} & \textbf{Case study size (max$|$variables$|$)} & \textbf{Verifying samples} & \textbf{Distribution/ diversity reported} \\\toprule
\cite{yuan1999modeling} & 1999 & 105 & Binary Decision Diagram & $\approx$1.3K & \noCircle & \noCircle \\
\cite{iyer2003race} & 2003 & 50 & Interval-propagation-based & 200 & \noCircle & \noCircle \\
\cite{yuan2004simplifying} & 2004 & 54 & Binary Decision Diagram & $<1$K & \noCircle & \noCircle \\
\cite{wei2004towards} & 2004 & 141 & Random Walk + {\sc WalkSat} & \multicolumn{3}{c}{\it No experiment conducted} \\
\cite{gogate2011samplesearch} & 2011 & 88 & Sampling via determinism & 6k & \noCircle & \noCircle \\
\cite{ermon2012uniform} & 2012 & 25 & {\sc MaxSat} + Search Tree &\multicolumn{3}{c}{\it Experiment details not reported} \\
\cite{chakraborty2014balancing} & 2014 & 29 & Hashing based & 400K & \noCircle & \halfCircle \\
\cite{chakraborty2015parallel} & 2015 & 28 & Hashing based (paralleling) & 400K & \noCircle & \halfCircle\\
\cite{meel2016constrained} & 2016 & 29 & Universal hashing & 400K & \noCircle & \halfCircle \\
\cite{dutra2018efficient} & 2018 & 5 & Z3 + \eq{one} flipping & 400K & \noCircle & \halfCircle \\\hline
\rowcolor{black!10}{\sc Snap} & 2020 & this   paper & Z3 + \eq{one} + local sampling & 400K & \yesCircle & \yesCircle\\
\bottomrule
\multicolumn{7}{r}{\noCircle~/ \yesCircle ~: the absence / presence of corresponding item \hspace{4pt} 
\halfCircle ~: only partial case studies \textit{(the small case studies)} were reported}\\
\end{tabular}
\end{center} 
\end{table*}
When developers extend   a code base, test suites let them  check that their new work does not harm old functionality.
Such  tests mean that developers can  find and fix more faults, sooner.
Hence, better tests enable faster code modification~\cite{fazlalizadeh2009prioritizing, lu2009introReg, haidry2013TCP}. 

By minimizing the number of tests  executed, we  also minimize the developer effort required to specify   the expected  behavior associated with
each test execution~\cite{yu2019terminator}.
If testing for (e.g.) core dumps, then specifying off-nominal behavior is trivial (just look for a core dump file).  But in many other  cases, specifying what should (and should not) be seen when a test executes  is a time-consuming task requiring a deep understanding of the purpose and context of the software.

Smaller tests suites are also cheaper to run.
The industrial experience is that excessive  testing can be   expensive and time consuming,
especially if   run after each modification to software~\cite{yu2019terminator}. 
Such high-frequency   testing can consumes as much as 80 percent of the software maintenance effort~\cite{chittimalli2007cost,yu2019terminator}.
Many current organizations spend tens of millions of dollars each year (or more) on
cloud-based facilities to  run large tests suites~\cite{yu2019terminator}.
The fewer the tests those
organizations have to run, the cheaper their testing.

Smaller test suites   are  also faster to execute. 
If minimal and effective test suites can be generated then,
  within a fixed time limit,   more faults can be found and fixed~\cite{yu2019terminator}.   Faster test execution means that software teams can certify a new release,  quicker.  
 This is particularly important for organizations using continuous integration since faster test suites mean they can make more releases each day -- which means that clients can sooner receive new (or fixed) features~\cite{parnin2017top}.

% For example, LexisNexis is an industrial company that provides legal research, risk management, and business research.  Due to market pressure,   LexisNexis wants to boast that their continuous integration practices allow them to    release multiple versions of their software, each day.  But some of their more complex software systems required   30 hours to execute~\cite{yu2019terminator}, thus impeding the desired   release velocity. Accordingly, LexisNexis asked this research team to find smaller test suites so they can:
% \begin{enumerate}
%     \item
%     Test software more often, then ship more updates   to customers, at a faster rate;
%     \item
%     Save   time when waiting for  user  feedback on     changes~\cite{yu2019terminator}.
% \end{enumerate}

\BLACK
% Despite all the above work, SE testing   via DPLL is still  intraotable; i.e. scales poorly   for larger problems  This is troubling since, as shown by the examples
% presented later in this paper, propositional formula extracted from software may have many variables. Hence, many researchers have turned to hybrid  methods that combine theorem provers with other methods (see below).

% We conjecture that finding backdoors can be done much quicker
% than the methods described above. Using the {\sc Snap} tactic,
% if many solutions share a few settings
% then all we need to is reason over the average values
% of a few randomly selected valid solutions.

% Before presenting that method, we first need to discuss how
% to implement software test generation using theorem provers (like a SAT
% solver).

\subsection{Theorem Proving in Software Engineering}\label{sect:problem}

As argued in this section,  the
test suite generation problem is so  complex that extra machinery
is needed to guide  theorem provers. 
The rest of this paper discusses
that extra machinery.

Many software problems   
can be  transformed into ``SAT''; i.e. a
propositional satisfiability problem.
For example, 
given a script of C programming, one can
translate it  into  CNF formulas, as done in  
\fig{codedemo}.
 Symbolic/dynamic execution techniques~\cite{baldoni2018survey,christakis2016guiding}  extract
 the possible execution branches of a procedural program.
 Each branch is
 a  conjunction of conditions $B_i=C_x \wedge C_y \wedge ...$
 so the whole program can be summarized as the disjunction
 $B_i \vee B_j \vee ...$. Using deMorgan's rules
 these clauses can be converted to
 conjunctive normal  form (CNF) where
 the inputs to the program are the variables in the CNF:
 \bi
 \item Disjunctions to conjunctions: $ P \vee  Q \equiv  (\neg P \wedge \neg Q)$
 \item Conjunctions to disjunctions:  $ \neg (P \wedge Q) \equiv \neg P \vee \neg Q$.
 \ei
 SAT   has applications in many areas, including test case generation.
Modern constraint solvers  (i.e. SAT-solvers) are based on some variant  of the   Davis-Putnam-Logemann-Loveland (DPLL) procedure~\cite{davis1960computing}.   DPLL  
searches systematically for a satisfying assignment, applying first unit propagation and pure literal elimination as often as possible. Then, DPLL branches on the truth value of a variable, and recurses.

Many methods have been explored to make these tools
 practical for large problems.
 For example 
Arito {\it et al.} proposed a framework to transform the test suite minimization problem (TSMP) in regression testing
into a constrained SAT problem~\cite{arito2012application}.
This transformation
is done by modeling TSMP instances as a set of Pseudo-Boolean constraints
that are later translated to SAT instances. 
 TSMP has two objectives: 1) minimizing the testing cost and 2) maximizing the program coverage.
To start with, a set of test cases $\mathcal{T}=\{t_1,t_2,t_3,\ldots\}$ as well as their 
running time cost $\{c_1,c_2,\ldots\}$ is defined. 
Here, 
$t_i$ is a binary signal indicating if the test case $i$ should be tested
and
the information about whether test case $t_i$ covers some element in the program $e_j$ is stored as the binary matrix $M=[m_{ij}]$.

To translate the TSMP into constrained problems, we use pseudo-boolean constraints: 
$\sum_{i=1}^n c_it_i\le B$ and $\sum_{j=1}^m e_j \ge P$
where $B\in \mathbb{Z}$ is the maximum allowed cost and  $P\in\{1,2,\ldots, m\}$ is the minimum coverage level.
Having the pseudo-boolean constraints, Een {\it et al.}~\cite{een2006translating} provides three techniques to translate pseudo-boolean constraints (linear constraints over boolean variables) into clauses that can be
handled by a SAT-solver.

% For the next steps, the pseudo boolean constraints provide big expressive power and could be translated to common constrained form in an automatic way~\cite{een2006translating}.

Combinatorial testing covers interactions of parameters in the system under test. A well-chosen sampling
mechanism can reduce the cost of software and system testing
by reducing the number of test cases to be executed~\cite{yamada2015optimization}.  
Note that   not all combinations are valid. For example,  MacOS does not support AMD processor while  IE does not support
MacOS, etc.
All of such constraints can be expressed as the feature model~\cite{janota2008model} or as product lines.
Further, such a feature model can be transformed into the CNF formulas~\cite{mendonca2009sat},
at which point,  SAT solvers can compute out the valid testing environment combination.
For other applications in this area, see~\cite{nie2011survey}.

In summary, in theory,  it is can be useful to reformulate SE tasks as a SAT task. As Micheal Lowry said at a panel at ASE'15:
\begin{quote}
{\em ``It used to be that reduction to SAT proved a problem's intractability.
But with the new     SAT solvers, that reduction now demonstrates
practicality.''}
\end{quote}
However, in practice,   general  SAT solvers, such as the Z3~\cite{de2008z3}, MathSAT~\cite{bruttomesso2008mathsat}, vZ~\cite{bjorner2015nuz} {\it et al.}, 
are challenged by the complexity of   real-world software models. For example, the 
largest benchmark for SAT Competition 2017~\cite{heule2013sat}
had 58,000 variables-- which is far smaller than (e.g.)
the   300,000 variable problems seen in the recent
SE testing literature~\cite{dutra2018efficient}.
Accordingly, the rest of this paper discusses ways to better customize theorem provers
in order to handle very large test case generation problems.

\subsection{ Theorem Prover For Large Problems}\label{sect:background}

As shown in Table~\ref{tbl:refs}, much prior research has explored scaling theorem proving for software engineering.
One way to tame the theorem proving problem is to simplify or decompose the CNF formulas. A recent example in this arena was  {\it GreenTire}, proposed by
Jia {\it et al.}~\cite{jia2015enhancing}.
{\it GreenTire} supports constraint reuse based on the logical implication relation among
constraints.
One advantage of this approach
is its efficiency guarantees. Similar to the analytical methods in linear programming, they are always applied to a specific class of problem.
However, even with the improved theorem prover,  such methods may be difficult to be adopted in large models.
{\it GreenTire} was tested in 7 case studies. Each case study was corresponding to a small code script with ten lines of code, e.g. the {\it BinTree} in \cite{visser2006test}.
For the larger models, such as those explored in this paper,
the following methods might do better.

Another approach,
which we will call
 sampling, is to combine
 theorem provers  Z3
  with stochastic sampling heuristics. 
  For example, given random selections for $b,c$, \eq{one} might be
  used to generate a new test suite, without calling a theorem prover.
  Theorem proving might then be applied to some (small) subset of the newly generated tests, just to assess how well the heuristics are working. 
  
% XXXX so note on scale. for large models. but the slow ness seen in into when qs use it
%. yuck. in this workm we use Z3 but ensure then we call it aminimum numner of times.
%if you want large models. nhundres of hurs. later infig7. show that we are orders of mangite slowe XXXX

The earliest sampling tools were based on binary decision diagrams (BDDs)~\cite{akers1978binary}.
Yuan {\it et al.}~\cite{yuan1999modeling, yuan2004simplifying} build
a BDD from the input constraint model and then weighted the branches of the vertices in the tree
such that a stochastic  walk from root to the leaf was
 able to generate samples with the desired distribution.
 In other work,
Iyer proposed a technique named {\it RACE} which has been applied in multiple industrial solutions ~\cite{iyer2003race}.  {\it RACE} (a)~builds a high-level model to represent the constraints; then (b)~implements a branch-and-bound algorithm for sampling diverse solutions.
The advantage of {\it RACE} is its implementation simplicity.
However,   {\it RACE}, as well as the BDD-based approached introduced above, return
highly biased samples, that is, highly non-uniform samples. For testing, this is not recommended since it means small parts of the code get explored at a much higher frequency than others.

Using a SAT solver {\it WalkSat}~\cite{selman1993local}, Wei {\it et al.}~\cite{wei2004towards} proposed {\it SampleSAT}. {\it SampleSAT} combines random walk steps with greedy
steps from {\it WalkSat}-- a method that works well for small  models.
However, due to the greedy nature of {\it WalkSat}, the performance of {\it SampleSAT} is highly
skewed as the size of the constraint model increases.
%In 2011, Gogate {\it et al.} move forward the frontier via  developing effective importance  sampling algorithms for mixed probabilistic and deterministic graphical models.

For seeking diverse samples, some use universal hashing~\cite{mansour1993computational}
which offers   strong guarantees of uniformity.
Meel {\it et al.}~\cite{meel2016constrained} list  key ingredients of integration of universal hashing and SAT solvers; e.g.  guarantee uniform solutions to a constraint model.
These hashing algorithms can be applied to the extreme large models (with near 0.5M variables).
%their work, was fouxes on a theoretial gaureenteee of diversity
%int ehroy, veryinteresyting. in practice, very slow. and as we sahall shall. generating a diviersity can be
% be ahcieved by many other methods include the random sampling of {\sc Snap}.
More recently, several improved hashing-based techniques have been purposed to balance the 
scalability of the algorithm as well as diversity (i.e. uniform distribution) requirements.
For example, Chakraborty {\it et al.} proposed an algorithm named {\it UniGen}~\cite{chakraborty2014balancing},
following by the {\it Unigen2}~\cite{chakraborty2015parallel}.
{\it UniGen} provides strong theoretical guarantees on the uniformity of generated solutions and has applied to constraint models with hundreds of thousands of variables.
However,   {\it UniGen}  suffered from  a large computation resource requirement. 
Later work explored a parallel version of this approach.
{\it Unigen2}   achieved near linear speedup on the number of CPU cores.
% {\em QuickSampler} out performsunigen2. and do better than {\sc Snap}. (measured on faster)
% the {\em QuickSampler} outperforms the Unigen2
% and sampling faster than {\sc Snap}, measuring by "how many samples generated in unit time"

To the best of our knowledge, the state-of-the-art technique for
generating test cases using theorem provers is  {\em QuickSampler}~\cite{dutra2018efficient}.
{\em QuickSampler} was evaluated on large real-world case studies, some of which have
more than 400K variables. At {\it ICSE'18}, it was shown that {\em QuickSampler} outperforms aforementioned {\it Unigen2} as well as another similar technique named {\it SearchTreeSampler}~\cite{ermon2012uniform}.
{\em QuickSampler} starts from a set of valid solutions generated by Z3. Next, it computes
the differences between the solutions using \eq{one}.
New test cases
generated in this manner are not guaranteed to be valid.  {\em QuickSampler} defines
three   terms, we use  later in this paper:
\bi
 \item A test suite is a set of valid tests.
 \item A test is {\em valid} if it uses input settings that satisfy the CNF.
 \item One test suite is more {\em diverse} than another if it uses more variable within the CNF disjunctions.  {\em Diverse} test suites are preferred since they cover more parts of the code.
\ei

According to Dutra {\it et al.}'s 
experiments,   the percent of valid
tests found by {\em QuickSampler}
 can be higher than 70\%. The percent of valid
 tests found by  {\sc Snap}, on the other hand, is 100\%.
 Further, as shown below, {\sc Snap} builds those tests with enough diversity much faster than
 {\em QuickSampler}.

\section{Implementing the {\sc Snap} Tactic  }\label{sect:aboutSnap}

% As stated in the introduction,  our approach combines
%  \eq{one} with 
% the Z3 with the {\sc Snap} tactic; i.e. we sample around the average values seen in a few 
% randomly selected valid tests.   

In the {\sc Snap} algorithm of \fig{frame},
each test is a set of  zeros or ones  (false, true) assigned to  all the variables in a CNF formula.
% Note that this algorithm  requires a {\em repair} function for step 3biv, and a  {\em termination} function for step 4a. Those two functions are  discussed in the next section.

As shown in     {\bf initial samples} (steps 1a,1b),
instead of computing some deltas between many
tests, {\sc Snap} restrains   mutation to  the deltas between a few   valid tests (generated from Z3).
{\sc Snap} builds a pool of  10,000 deltas from $N=100$ valid tests
(which mean   calling  a theorem prover  only  $N=100$ times).
{\sc Snap}  uses this pool as a set of candidate ``mutators''
for existing tests (and by ``mutator'', we mean an operation
that converts an existing test into a new one).

After that, in     {\bf delta preparation} (steps 2a,2b),  {\sc Snap} applies \eq{one}.
Note that the more often a setting repeats, the more likely it is a backdoor variable.
Hence, step 2b sorts the deltas on occurrence frequency.  This sort is used in step 3b.

In {\bf sample} (steps 3a,3b), {\sc Snaps} samples around the average  values seen in a few randomly selected valid tests. Here, "averaging" is inferred by using the median values
seen in $k$ clusters.    Note that, in step 3b, we use deltas that are more likely
to be valid (i.e. we use the deltas that occur more frequently).

Step 3b.iii  is where we verify the new candidate using
Z3. {\sc Snap} explores far fewer candidates than {\em QuickSampler} (10 to 750 times less, see \tion{howmany}). Since we are exploring less, we can take the time to verify them all.
Hence, 100\% of {\sc Snap}'s tests are valid (and the same is {\em not} true for {\em QuickSampler}-- see \fig{hit}).

Note that in  3b.iv, we only add our new tests to the  clusters if it fails
verification (taking care to first   repair it).
We do this since test cases that pass verification do not add new information. But   when an instance fails verification and is repaired,
that offers new settings.

Note also that {\sc Snap} takes great care in how it calls a theorem prover.
Theorem provers are  much slower for 
 {\em generating} new tests than {\em repairing} invalid tests
 than for {\em verifying} that a test is valid
 (since there are  more options for
 generation than for repairing than
 for verification). Hence, {\sc Snap} needs to
 {\em verify} more than it {\em repairs}
 (and also do {\em repairs} more than {\em generating} new tests).  
% The {\sc Snap} is based on the assumption that 
% \begin{center}
% \textit{
% The delta $\delta$ of two valid solution $a,b$ can be
% transferred (or grafted) into another valid solution $c$.}
% \end{center}
% As introduced in \fig{codedemo}, a valid solution in theorem proving is a binary vector e.g. ``11001...'', with same length as number
% of  variables in the CNF.
% The delta $\delta$ in this paper is defined as
% \begin{equation}\label{eq:delta}
% \delta(a,b) = a \oplus b
% \end{equation},
% where $a, b$ are two binary vectors, and $\oplus$ is the XOR (exclusive-or) for binary vectors.
% With these notations, the assumption can be parsed as 
% \begin{center}\it % quicj samoer assumption
% $c \oplus \delta(a,b)$ is \underline{{\it likely}} to be valid if $a,b,c$ are valid solutions.
% \end{center}
% %XXXX surrogate. we use small intiial sample
% % begineering transfer . {\sc Snap} reasons in local regin=on. verification partial 
% We will explore this assumption later in the experiment \tion{exp}.
More specifically:
\bi
\item The call to Z3  in step 1a is a {\em generating call}. This can be slowest since this  must navigate   all the constraints of our CNF. 
Therefore, we  only do this $N=100$ times.
\item
The call to Z3 in  step 3b.iii is a {\em verification call} and is much faster
since all the variables are set.
\item  
The call to Z3 in   step 3b.iv {\em repair} call, is  slower 
than step 3b.iii since   our repair   method adds
  open choices to a test.
\ei
Note that we only need to repair the small minority of new tests that fail verification.
Later in this paper, we can use \fig{hit} to show that
repairs are only needed on 30\% (median) of
all tests.

\subsection{Implementing ``Repair''} \label{sect:repair}

 {\sc Snap}'s repair function
deletes   ``dubious'' parts of a test case,
then uses Z3 to fill in the the gaps. In this way, when
we repair a test, most   bits are set and Z3 only has to
search a small space.

To find the ``dubious'' section, we reflect on how step 3b.ii operates.
Recall that the new test  uses
$ \delta = a \oplus b$ and $a,b$ are valid tests taken from {\em samples}.
Since $a,b$ were valid, then the ``dubious'' parts of
the test is anything that was not seen in both $a$ and $b$.
Hence, we preserve the bits in  $c\oplus\delta$ bits (where the corresponding $\delta$ bit was 1), while removing
all other bits (where $\delta$ bit was 0). For example:
\bi
\item
To  mutating  $c=$(1,0,0,1,1,0,0,0) use $\delta=$(1,0,1,0,
1,0,1,0). 
\item
If $c\oplus\delta=$(0,0,1,1,0,0,1,0) is invalid, then {\sc Snap}  deletes the ``dubious'' sections
as follows.
\item
{\sc Snap} preserves any ``1'' bits that were seen in $\delta$.
\item
{\sc Snap} deletes the others; e.g.  bits 2, 4, 6, 8 
(0,\xcancel{0},1,\xcancel{1},0,\xcancel{0},
1,\xcancel{0}).
\item
 Z3 is then called to figure out the missing bits of $(0?1?0?1?)$. 
 \ei
%   \colorbox{YellowOrange}{{\em Heuristic \#5}}
%   (shown above) is based on the assumption that 
%  these last step (where Z3 repairs the vector)
% is usually faster than generating
% a completely new solution from scratch.

% \newcommand{\noh}[1]{{\colorbox{white}#1}}
% \newcommand{\hhh}[1]{{\colorbox{black!10}#1}}
  
% \begin{align*}
% {\color{OliveGreen}\mathbf c} &=& 
% \noh{1}\noh{0}\noh{0}\noh{1}\noh{1}\noh{0}\noh{0}\noh{0} \\
% \delta &=& \hhh{1}\noh{0}\hhh{1}\noh{0}\hhh{1}\noh{0}\hhh{1}\noh{0} \\
% {\color{red} \mathbf c\oplus\delta} &=& \noh{0}\noh{0}\noh{1}\noh{1}\noh{0}\noh{0}\noh{1}\noh{0} \\\nonumber
% c\oplus(\delta\vee\delta'_{\mathit{z3-search}}) &=& \hhh{0}\noh{?}\hhh{1}\noh{?}\hhh{0}\noh{?}\hhh{1}\noh{?}\nonumber
% \end{align*} 

\begin{figure}[!t]
% \hline 
% \vspace{2mm}
 
 \begin{center}\fcolorbox{black}{black!5}{\begin{minipage}{\linewidth} 
 \be[start=0,label=\arabic*)]
 \item \textbf{Set up}
 \be
\item Let $N=100$; i.e. initial sample size;
\item Let $k=5$; i.e. number of clusters;
\item Let {\it suite}$=\emptyset$; i.e. the output test suite;
\item Let {\it samples}$=\emptyset$; i.e. a temporary work space.
\ee
\item \textbf{ Initial samples generation}:
    \be
    \item  Add $N$ solutions (from   Z3) to   {\it samples}
    \item Put all {\it samples} into $\mathit{suite}$ (since they are valid)
    \ee
\item \textbf{Delta  preparation}:
    \be
    \item Find delta $\delta=(a \oplus b)$ for all $a,b\in \mathit{samples}$. 
    \item Weight each delta by how often it repeats
    \ee
\item \textbf{Sampling}
    \be
    \item Find $k$ centroids in    {\it samples} using $k$-means ;
    \item For each centroid  $c$, repeat  $N$ times:
    \be
    \item    stochastically pick
      deltas $\delta_i$, $\delta_j$ at prob. equal to their weight. 
    \item compute  a new candidate using $c\oplus(\delta_i\vee
    \delta_j)$
    \item verify new  candidate using Z3;
    \item if invalid,  repair using Z3 (see \tion{repair}). Add to {\it sample};
    \item add the repaired candidate to  {\it suite};
    \ee
    \ee
\item {\bf Loop or terminate}:
\be
\item If improving  (see \tion{term}), go to step 2. Else return {\it suite}.

\ee
\ee
% \hline
\end{minipage}}\end{center}
\caption{{\sc Snap}}
\label{fig:frame}
\end{figure}
\subsection{Implementing ``Termination''}\label{sect:term}

To implement {\sc Snap}'s termination criteria (step 4a), we need
a working measure of diversity.
Recall from the introduction
that one  test  suite  is  more
{\em diverse}
than  another  if  it  uses
more  of  the  variable  settings  with  disjunctions  inside  the
CNF.
{\em Diverse}
test  suites  are
{\em better}
since  they  cover
more parts of the code.

To measure diversity, we used Feldt {\it et al.}~\cite{feldt2016test}'s   normalized compression distance (NCD). A test suite with high NCD implies   higher
code coverage during the testing\footnote{Aside:
we note that we did not adopt the diversity metric (distribution of samples displayed as a histogram) from~\cite{chakraborty2015parallel, dutra2018efficient} 
since computing that metric is very time-consuming. For the case
studies of this paper, that calculation required days of CPU.
Later in this paper, we show that our
use of this diversity measure is not a threat to validity for this study. }.  
%  NCD is based on information theory --
% the
% Kolmogorov complexity~\cite{li2013introduction} of a binary string $x$ is the length of 
% the shortest program that outputs $x$.
% NCD is based on the observation that the degree to which a string can be compressed by real-world compression programs
% (such as \texttt{gzip} or \texttt{bzip2}).
NCD uses \texttt{gzip} to the estimate  
Kolmogorov complexity ~\cite{li2013introduction} of the tests. 
If $C(x)$ is the length of   compression of $x$ and $C(X)$ is the compression length of   binary string set $X$'s concatenation, then:
\begin{equation}\label{eq:ncd}
\text{NCD}(X) = \frac{C(X)-\min_{x\in X}\{C(x)\}}{\max_{x\in X}\{C(X\backslash\{x\})\}}
\end{equation}
% \bi
% \item the histogram shows the occurrence of some solution patterns, which is not suitable in RQ2 analysis, since the goal of {\sc Snap} is not regarding counting. In fact, {\sc Snap} did not generate millions, or even tens of millions of samples as {\em QuickSampler} did.
% \item the histogram did not apply to all case studies, especially the large case studies. In~\cite{dutra2018efficient}, they only illustrated the results in the case studies for which number of samples generated by 
% Unigen2 was at least five time of total number of solutions. Among 30 case studies, only five of them were reported.
% \item the calculation of that metric is time-consuming. It would be approximately 20 CPU days.
% \ei
% {\sc Snap}  exits if    NCD improves by  
% $X\le 5\%$ in the last $T=10$ minutes.

% XXX @timm this is the revision at round 1
\BLUE
To understand how the NCD is revealing the diversity of a test suite, consider the following test suite {where each row in the matrix represents one test case}:
\newcommand{\TS}{\colorbox{red!35}{T1}}
\newcommand{\TSS}{\colorbox{orange!25}{T2}}
\newcommand{\TSSS}{\colorbox{yellow!10}{T3}}
\begin{equation*}
 \TS = \begin{bmatrix}
0 & 0 & 0 & 0 & 0 &\\
0 & 0 & 0 & 1 & 1 &\\
0 & 0 & 0 & 1 & 0 & 
\end{bmatrix}
\end{equation*}
Here,
NCD$_1 = 0.142$.
Now,
assuming that we have obtained the following test suite after several iterations
\begin{equation*}
\TSS = \begin{bmatrix}
0 & 0 & 0 & 0 & 0 & \\
0 & 0 & 0 & 1 & 1 & \\
0 & 0 & 0 & 1 & 0 &\\
1 & 0 & 0 & 0 & 0 & \colorbox{green!20}{$+$} \\
0 & 1 & 0 & 1 & 1 & \colorbox{green!20}{$+$}  \\
\end{bmatrix}
\end{equation*}
then NCD$_2$  is now 0.272.
Note that (a)~\colorbox{green!20}{$+$} 
marks the new test cases obtained since \TS;
and (b)~NCD$_2$ is larger since the new cases  cover various options in first two bits.

On the other hand, if we further consider the following test suite:

\begin{equation*}
\TSSS = \begin{bmatrix}
0 & 0 & 0 & 0 & 0 & \\
0 & 0 & 0 & 1 & 1 & \\
0 & 0 & 0 & 1 & 0 &\\
0 & 0 & 0 & 0 & 1 & \colorbox{green!20}{$+$} \\
1 & 0 & 0 & 0 & 0 &\\
0 & 1 & 0 & 1 & 1 & \\
0 & 1 & 0 & 1 & 0 & \colorbox{green!20}{$+$} 
\end{bmatrix}
\end{equation*} then NCD$_3$ = 0.305.
Here, due to two new test cases (marked as  \colorbox{green!20}{$+$}), we do see NCD improvements
in \TSSS, as compared to \TSS.
Such scale of improvements, however, is not significant: from \TS to \TSS, we got $\mathit{\frac{0.272-0.142}{0.142}=}91\%$ NCD improvements, while in the \TSSS, we got $\mathit{\frac{0.305-0.272}{0.272}=}12.1\%$ increases. This is because the new
cases in \TSSS~ does not explore the diversity of new bits, such as the third bit.
\footnote{In this example, we examine the
diversity via single bit.
However, the NCD also examines the bits-group, such as the combinations of bit pair $(x,y)$, or bit tuple $(x,y,z)$ etc.
}

Another   point
to note 
is that NCD is presenting the diversity of a string-block
(i.e. every substring matters).
One substring $x$ where $C(x)\to 0$ does not imply NCD$\to 1$.
This is because the $x$ itself can attribute a lot to NCD of the whole string-block. Take the aforementioned \TS~ as an example: among all three cases $x_1$, $x_2$ and $x_3$, $C(x_1) \to 1$,  NCD$(x_1\cup x_2\cup x_3) \ll 1$.

\BLACK

\subsection{Engineering Choices}\label{sect:choice}
{\sc Snap} uses theses
control parameters:
\bi
\item $X=5\%$;
\item $T=10$ minutes;
\item $N=100$ samples;
\item $k=5$ clusters.
\ei
In future work, it could be insightful to vary these values.
Another area that might bear further investigation is the clustering method used in step 3a.  For this paper, we tried different clustering methods. Clustering ran so fast that we were not motivated to explore alternate algorithms. Also, we  found that the details of the clustering were less important than pruning away
most of the items within each cluster (so that we only mutate the centroid).

\section{Experimental Set-up}\label{sect:exp}

\subsection{Code}\label{sect:code}
To explore the research questions shown in the introduction,
the {\sc Snap} system shown in \fig{frame} 
was implemented in C++
using    Z3 v4.8.4 (the latest release when the experiment was conducted).
A $k$-means cluster was added using the free edition of ALGLIB~\cite{bochkanov2013alglib}, a numerical analysis and data processing library delivered for free under GPL or Personal/Academic license.
{\em QuickSampler} does not integrate the samples verification into the workflow. Hence, in the experiment, we adjusted the workflow of {\em QuickSampler}
so that all samples are verified before termination, which is the same as {\sc Snap} as in \tion{term}.
Also, the outputs of {\em QuickSampler} were the assignments of independent support.
The {\em independent support} is a subset of variables which completely determines all the assignments to a formula~\cite{dutra2018efficient}.
In practice, engineers need the complete test case input;
consequently, for valid samples, we extended the {\em QuickSampler} to get full assignments of all variables from independent support's assignment via propagation.

% Since a  software engineer, he or she wants to get the full assignment of all variables.  Consequently, for valid samples, we extended the {\em QuickSampler} to get full assignment from independent support's assignment via propagation.

\newcommand{\tm}{\rowcolor{blue!10}}
\newcommand{\tl}{\rowcolor{orange!10}}
\begin{table} 
\caption{Case studies used in this paper. Sorted by number of variables. Medium sized-problems are highlighted with {\color{blue} blue rows} while the large ones are in {\color{orange} orange rows}. Three items (marked with *) are not included in some further reports (see text).
See \tion{case_studies} for details.}
\label{tab:benchmarks}
% \footnotesize
 \small
\begin{center}
% \begin{tabular}{p{1in}|lp{1in}p{1in}}
\begin{tabular}{c|lll}
\hline
\rowcolor{black!20}\textbf{Size}&\textbf{Case studies} & \textbf{Vars} & \textbf{Clauses}  \\\hline
&blasted\_case47 & 118 & 328 \\
&blasted\_case110 & 287 & 1263 \\
&s820a\_7\_4 & 616 & 1703 \\
&s820a\_15\_7 & 685 & 1987 \\
&s1238a\_3\_2 & 685 & 1850 \\
Small &s1196a\_3\_2 & 689 & 1805 \\
&s832a\_15\_7 & 693 & 2017 \\
&blasted\_case\_1\_b12\_2* & 827 & 2725 \\
&blasted\_squaring16* & 1627 & 5835 \\
&blasted\_squaring7* & 1628 & 5837 \\
&70.sk\_3\_40 & 4669 & 15864 \\
&ProcessBean.sk\_8\_64 & 4767 & 14458 \\
&56.sk\_6\_38 & 4836 & 17828 \\
&35.sk\_3\_52 & 4894 & 10547 \\
&80.sk\_2\_48 & 4963 & 17060 \\\hline
\tm & 7.sk\_4\_50 & 6674 & 24816 \\
\tm & doublyLinkedList.sk\_8\_37 & 6889 & 26918 \\
\tm & 19.sk\_3\_48 & 6984 & 23867 \\
\tm & 29.sk\_3\_45 & 8857 & 31557 \\
\tm Medium& isolateRightmost.sk\_7\_481 & 10024 & 35275 \\
\tm & 17.sk\_3\_45 & 10081 & 27056 \\
\tm & 81.sk\_5\_51 & 10764 & 38006 \\
\tm & LoginService2.sk\_23\_36 & 11510 & 41411 \\\hline
\tl & sort.sk\_8\_52 & 12124 & 49611 \\
\tl & parity.sk\_11\_11 & 13115 & 47506 \\
\tl & 77.sk\_3\_44 & 14524 & 27573 \\
\tl Large& 20.sk\_1\_51 & 15465 & 60994 \\
\tl & enqueueSeqSK.sk\_10\_42 & 16465 & 58515 \\
\tl & karatsuba.sk\_7\_41 & 19593 & 82417 \\
\tl & tutorial3.sk\_4\_31 & 486193 & 2598178 \\
\hline
\end{tabular}
\end{center}
\end{table}

\begin{figure*}[!b]
\centering
\includegraphics[width=0.85\textwidth]{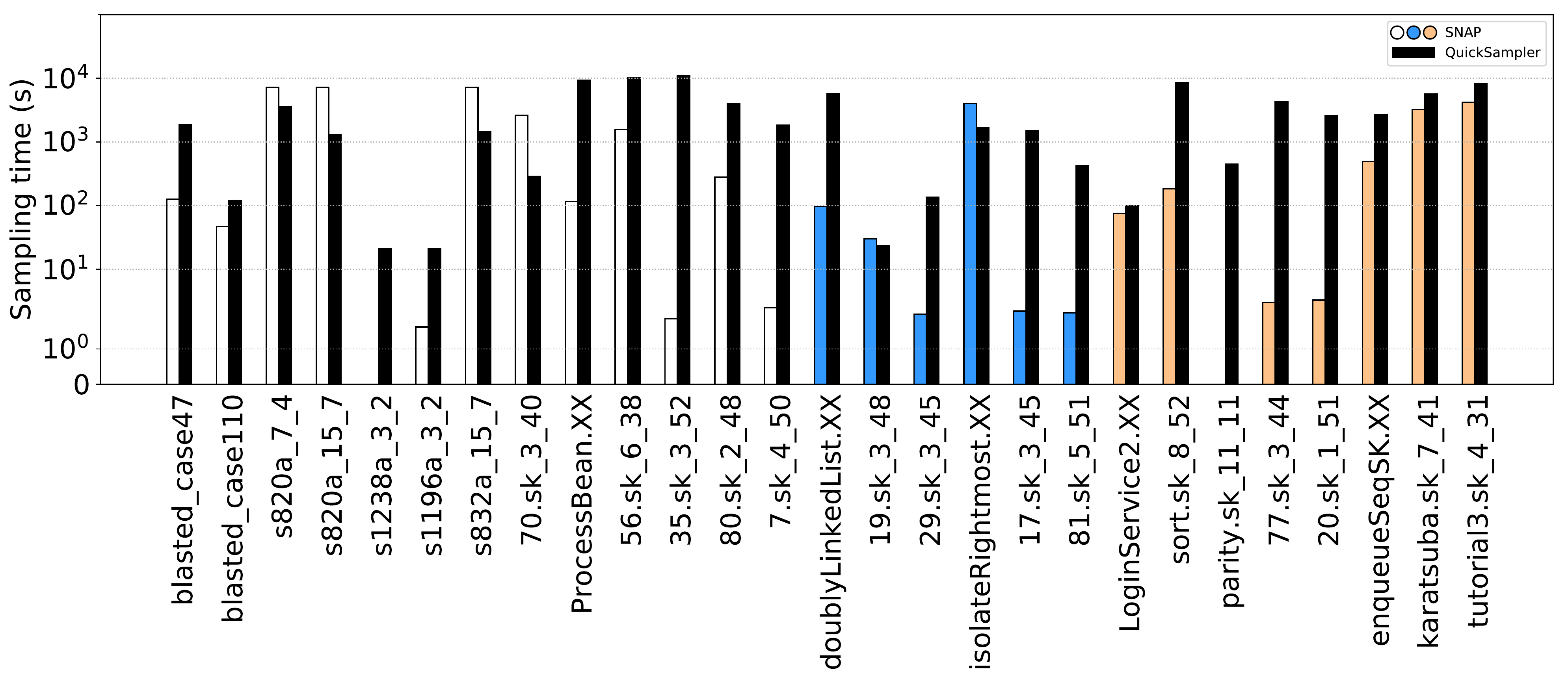}
\caption{RQ1 results: Time to  terminated (seconds),
The y-axis is in log scale.
The {\sc Snap} sampling time for {\it s1238\_a\_3\_2} and {\it parity.sk\_11\_11} is not reported since their achieved NCD were much worse than {\em QuickSampler}'s (see \fig{rq21}). \fig{rq_speedup} illustrates the corresponding speedups.}
\label{fig:rq22}
\end{figure*}

\subsection{Case Studies}\label{sect:case_studies}

\tab{benchmarks} lists the case studies used in this work.
We can see that the number of variables ranges from hundreds to more than 486K.
The large examples have more than 50K clauses, which is very huge.
For exposition purposes,  we divided the case studies into three groups: the small case studies with vars $< 6K$; the medium case studies with $6K < $ vars $ < 12K$ and the large case studies with vars $> 12K$.  

For the following reasons,
our case studies are the same as those used in   the 
{\em QuickSampler} paper:
\bi
\item We wanted to compare our method to {\em QuickSampler};
\item Their case studies were online available; 
\item Their   studies are used
in many papers~\cite{chakraborty2014balancing,chakraborty2015parallel,meel2016constrained,dutra2018efficient}.
\ei
These case studies are representative of scenarios engineers met in software testing or circuit testing in embedded system design. 
They include bit-blasted versions of SMTLib case studies, ISCAS89 circuits augmented with parity conditions on randomly chosen subsets of outputs and next-state variables,
problems arising from automated program synthesis and constraints arising in bounded theorem proving. 
For more introduction of the case studies, please see ~\cite{chakraborty2015parallel,dutra2018efficient}.

% Given the CPU running times, one can simply imply that we cannot explore large amount of solutions in theorem proving only
% with  SAT solver.
% While on the other hand, without the help of Z3 solvers, starting from zero knowledge, 
% seeking for valid solutions is also challenging, given the highly constraint models. 
% Therefore, with the help of Z3 solvers, generating more samples via some technique is 
% a tenable idea.

For pragmatic reasons, certain case studies were omitted from our study.
For example, we do not report on {\it diagStencilClean.sk\_41\_36} in the experiment since the purpose of this paper is to sample a set of valid solutions to meet the diversity requirement; while there are only 13 valid solutions from this model. The {\em QuickSampler} spent 20 minutes (on average) to
search for one solution.

Also, we do report on the case studies marked
with a  star(*) in \tab{benchmarks}.
Based on the experiment, we found that even though the {\em QuickSampler} generates
tens of millions of samples for these examples, all samples were the assignment to the {\em independent support} (defined in \tion{code}). 
The omission of these case studies is not a critical issue.
Solving or sampling these examples is not difficult; since they are all very small, as compared to other
larger case studies.

\subsection{Experimental Rig}

We compared {\sc Snap} to  the state-of-the-art   {\em QuickSampler}, technique.
To ensure a repeatability,  
we update the Z3 solver in {\em QuickSampler} to the latest version.
  
To reduce the observation error and test the performance robustness,
we repeated all experiment 30 times with 30 different random seeds.
To simulate real practice, such random seeds were used in Z3 solver (for initial solution generation), ALGLIB (for the $k$-means) and other components.
Due to   space limitation, we cannot report results for all 30 repeats. Instead,
we report the medium or the IQR (75-25th variations) results.

All experiments were conducted on Xeon-E5@2GHz machines with 4GB memory, running CentOS. 
We only used one core per machine.

\section{Results}\label{sect:results}

The rest of this paper use the machinery defined above to answer the four
research questions posed in the introduction.

% Before answering those questions, we offer one  observation
% seen when preparing
% to answer our research
% questions.
% \tab{benchmarks} color-coded our case studies such that the medium to large case studies are shown in blue and orange. Looking across all the following results, with
% the exception of runtimes, we see no pattern in the color coding (e.g. it is {\em not} true that larger problems have more duplicates).  We tried various data mining methods to learn predictors
% for slow/fast test generation but
% failed to find convincing results.
% This failure means that the only
% way we know (so far) to check if
% a program's test case is slow to generate is to actually try to generate those tests.
% The complexity of test suite generation comes from the relationships between the variables, and not necessarily the number of variables themselves.

\subsection{ RQ1: How Much Faster is the {\sc Snap} Tactic? }

 \fig{rq22} shows the execution time required for {\sc Snap}
 and {\em QuickSampler}. The y-axis of this plot is a log-scale
 and shows time in seconds.
 These results
 are shown in the same order as \tab{benchmarks}. That is, from left to right, these case studies
grow from around 300 to around 3,000,000 clauses.

For the smaller case studies, shown on the left, {\sc Snap} is sometimes slower
than {\em QuickSampler}. Moving left to right, from smaller to larger case studies,
it can be seen that {\sc Snap} often terminates much faster than  {\em QuickSampler}.
On the very right-hand side of \fig{rq22}, there are some results where it seems {\sc Snap} is not particularly fastest. This is due to the log-scale applied to the y-axis. Even in these cases, {\sc Snap} is terminating in less than an hour while other approaches need more than two hours.

\begin{figure}[!t]
\begin{center}
\includegraphics[width=0.48\textwidth]{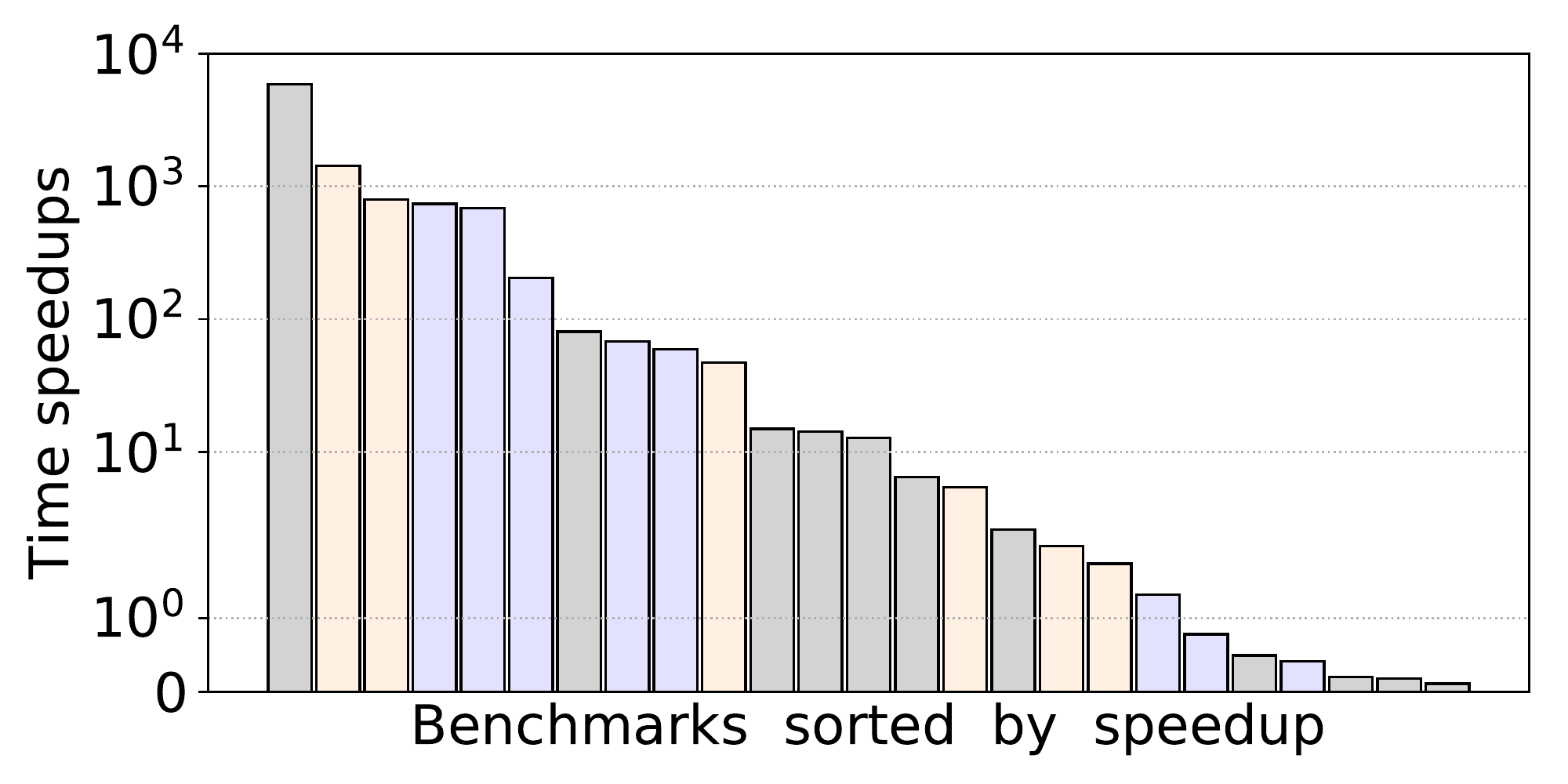}
\end{center}
\caption{RQ1 results: sorted $\mathit{speedup}$ (time({\em QuickSampler}) / time({\sc Snap})).
If over $10^0$, then  {\sc Snap} terminates earlier.}
\label{fig:rq_speedup}
\end{figure}
\fig{rq_speedup} is a summary of  \fig{rq22} that divides the execution time for both systems.  From this figure it can be seen:

\begin{result}{1}
~{\sc Snap}
terminated 10 to 3000 times faster
than {\em QuickSampler} (median to max).
\end{result}

There are some exceptions to this conclusion, where {\em QuickSampler}
was faster than {\sc Snap} (see the right-hand-side of \fig{rq_speedup}).
Those cases are usually for small models  (17,000 clauses
or less). For medium to larger models, with 20,000 to 2.5 million clauses,
{\sc Snap} is often orders of magnitude faster.

\begin{table} 
% \footnotesize
%\small
\caption{RQ2 results: number of unique valid cases in   test suite. Sorted
by last column. Same color scheme as \tab{benchmarks}. 
}
\begin{center}
% \begin{tabular}{lp{1in}p{1.5in}|p{1in}}
\begin{tabular}{lrr|r}
\toprule
 &$\mathbf{S_S}$   & 
$\mathbf{S_Q}$   &  $\mathbf{S_Q}/$\\
{\bf Case studies} &   \bfseries{\sc{{\sc Snap}}} & 
{ {\em QuickSampler}} & $\mathbf{S_S}$  \\\toprule
blasted\_case47 & 2899 & 71 & 0.02 \\

\tm isolateRightmost & 15480 & 7510 & 0.49 \\
\tl LoginService2 & 404 & 210 & 0.52 \\
\tm 19.sk\_3\_48 & 204 & 200 & 0.98 \\
70.sk\_3\_40 & 3050 & 4270 & 1.40 \\
s820a\_15\_7 & 29065 & 70099 & 2.41 \\
\tm 29.sk\_3\_45 & 225 & 660 & 2.93 \\
s820a\_7\_4 & 37463 & 124457 & 3.32 \\
s832a\_15\_7 & 27540 & 96764 & 3.51 \\
s1196a\_3\_2 & 225 & 1890 & 8.40 \\
\tl enqueueSeqSK & 338 & 2495 & 7.38 \\
blasted\_case110 & 274 & 2386 & 8.71 \\
\tl tutorial3.sk\_4\_31 & 336 & 2953 & 8.79\\
\tm 81.sk\_5\_51 & 227 & 2814 & 12.40 \\
\tl sort.sk\_8\_52 & 812 & 10184 & 12.54 \\
\tl karatsuba.sk\_7\_41 & 139 & 4210 & 30.29 \\
\tl 20.sk\_1\_51 & 239 & 10039 & 42.00 \\
\tm doublyLinkedList & 278 & 12042 & 43.32 \\
\tm 17.sk\_3\_45 & 228 & 12780 & 56.05 \\
ProcessBean & 1193 & 75392 & 63.20 \\
\tm 7.sk\_4\_50 & 258 & 18090 & 70.12 \\
56.sk\_6\_38 & 1827 & 149031 & 81.57 \\
80.sk\_2\_48 & 653 & 54440 & 83.37 \\
\tl 77.sk\_3\_44 & 245 & 33858 & 138.20 \\
35.sk\_3\_52 & 258 & 193920 & 751.63 \\
\bottomrule
\end{tabular}
\end{center}

\label{tab:rq4}
\end{table}

\begin{figure} 
\begin{center}
\centering
\includegraphics[width=0.38\textwidth]{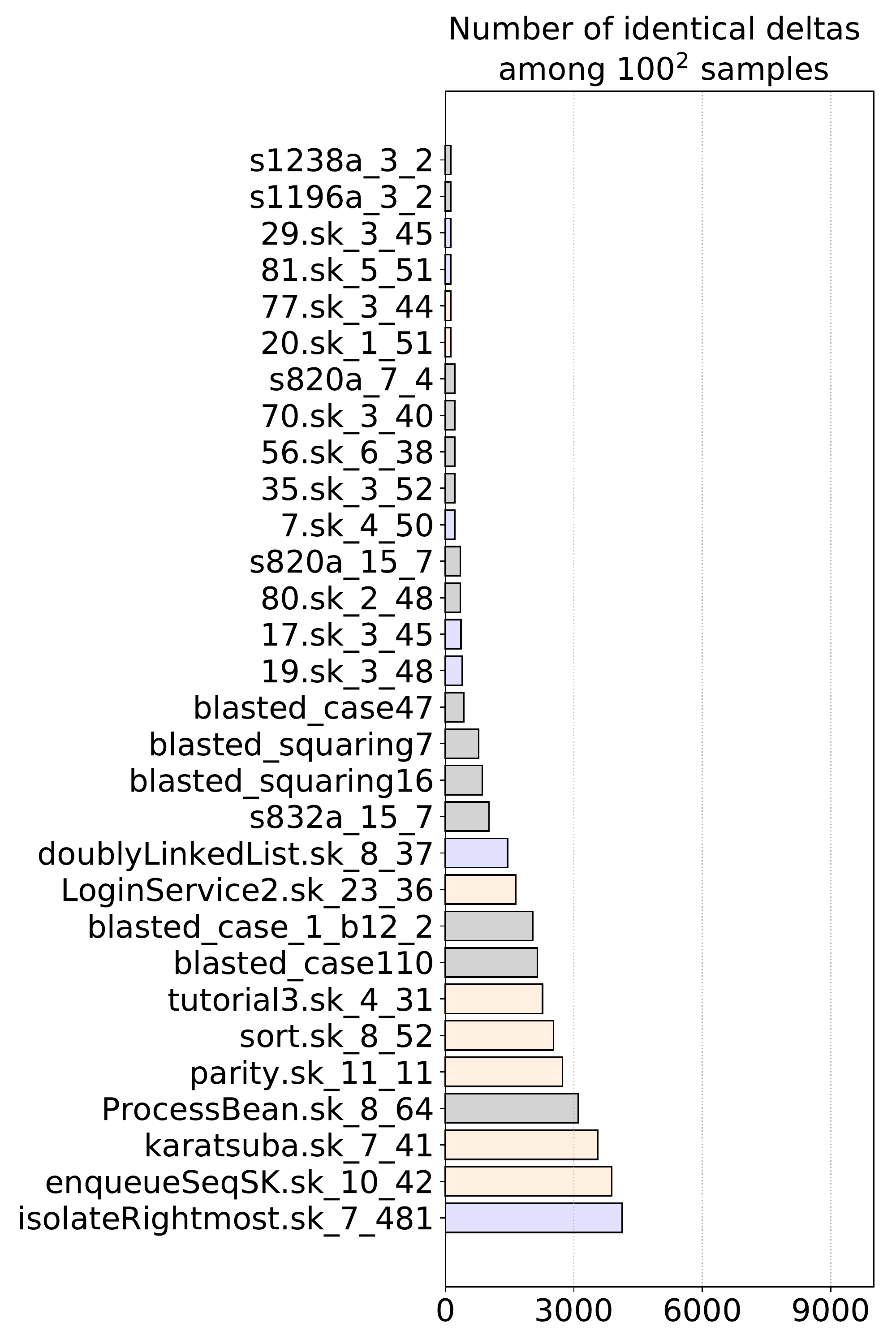}
\end{center}
\caption{ Identical deltas seen in 100*100 pair of valid solution deltas for all case studies. Same color scheme as \tab{benchmarks}.}
\label{fig:delta_iden}
\end{figure}

\subsection{RQ2: Does  the {\sc Snap} tactic find  fewer  test  cases?}\label{sect:howmany}

\tab{rq4} compares the number of tests 
from {\em QuickSampler} and {\sc Snap}. As shown by the last column in that table:
\begin{result}{2}
Test  cases  from  {\sc Snap} were  10  to  750 times smaller than from {\em QuickSampler} (median to max).
\end{result}
Hence we say that using {\sc Snap} is easier than other methods, where  ``easier'' is defined as per our {\em Introduction}. That is, when test suites are 10 to 750 times smaller,
then they are faster to run,
consumes less cloud-compute resources, and means developers have to spend less time processing failed tests.

\subsection{RQ3: How ``good'' are the tests found via the {\sc Snap} tactic? }

Generating small tests sets, and doing so very quickly,
is not interesting {\em unless}
those test suites are also ``good''. 
This section applies two definitions of ``good''
to the {\sc Snap} output:
\bi
\item \BLUE
{\em Credibility:} Recalling \fig{codedemo},
test suites need to satisfy the CNF clauses generated from source code.
As defined in the introduction, we say that
the ``more credible'' a test suite, the larger the percentage of valid tests.
\BLACK
\item {\em Diversity:} A CNF clause is conjunction of disjunctions.
Diversity measures how many of   disjunctions are explored by
the tests. This is important since a high diversity means
that most code branches   are covered. 
\ei

\BLUE
\subsubsection{Credibility}\label{sect:validity}

Regarding {\em credibility}, we note that   {\sc Snap}  
only prints valid tests. That is, 100\% of {\sc Snap}'s tests
are valid.

\BLACK

The same cannot be said for  {\em QuickSampler}.
That algorithm ran  so quickly since it assumed that tests generated
using   \eq{one} did not need verification. To check
that assumption,
for each case study, we randomly generated 100 valid solutions, $S=\{s_1,s_2,\ldots s_{100}\}$ using
Z3. Next, we selected three $\{a,b,c\} \in Ss$ and built  a new test case
using \eq{one}; i.e. $\mathit{new}=c\oplus(a \oplus b)$. 
 
\fig{delta_iden} lists  the number of identical deltas seen in $100^2$   of those deltas.
We rarely found large sets of unique deltas; i.e. 
  among the 100 valid solutions given by Z3,
many  $\delta$s were shared within pairwise solutions.
This is important since if otherwise, the \eq{one} heuristic would be dubious.
% This is important since,
% if otherwise, the \eq{one} does not hold.
 
\begin{figure}[!t]
\begin{center}
\centering
\includegraphics[width=0.4\textwidth]{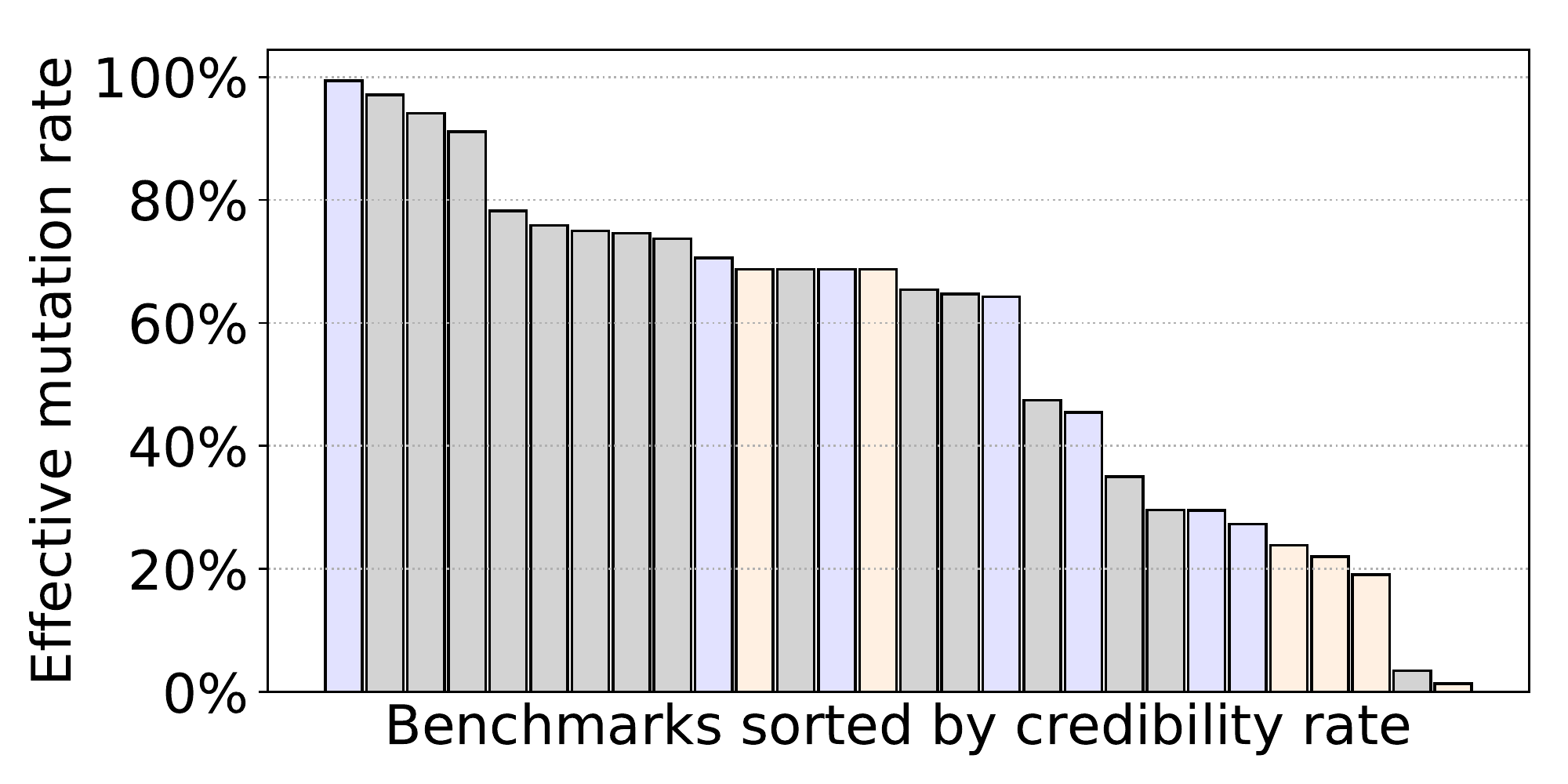}
\end{center}
\caption{RQ3 results for ``credibility'': percentage of valid mutations found it step3b.iii
(computed separately for each case study).}
\label{fig:hit}
\end{figure}

The percentage of these deltas that proved to be valid in step3b.iii of Algorithm 1 are shown in \fig{hit}.
 Dutra {\it et al.}'s estimate was that   the percentage of valid tests generated by
 \eq{one} was
 usually 70\% or more. As shown by the median values of  \fig{hit}, this was indeed the case. However, we
 also see that in the lower third of those results,
 the percent
of valid tests generated by \eq{one} is very low: 25\% to 50\% 
(median to max). 
 
This result   make us cautious about
using {\em QuickSampler} since, when the \eq{one} heuristics fails, it seems to be inefficient.

By way of comparisons, it is relevant
to remark here that  
{\sc Snap} verifies every test case it generates. This is practical
for {\sc Snap}, but impractical for {\em QuickSampler} since these
two systems typically process $10^2$ to $10^8$ test cases, respectively.
In any case, another reason to recommend {\sc Snap} 
is that this tool delivers tests suites where 100\% of all tests are valid.

In summary, measured in terms of {\em credibility}:
\bi
\item
{\sc Snap}'s tests are 100\% ``good'' 
\item While  other methods may find fewer   ``good'' tests.
\ei

\subsubsection{Diversity}

\begin{figure}[!t]
\begin{center}
 \includegraphics[width=0.38\textwidth]{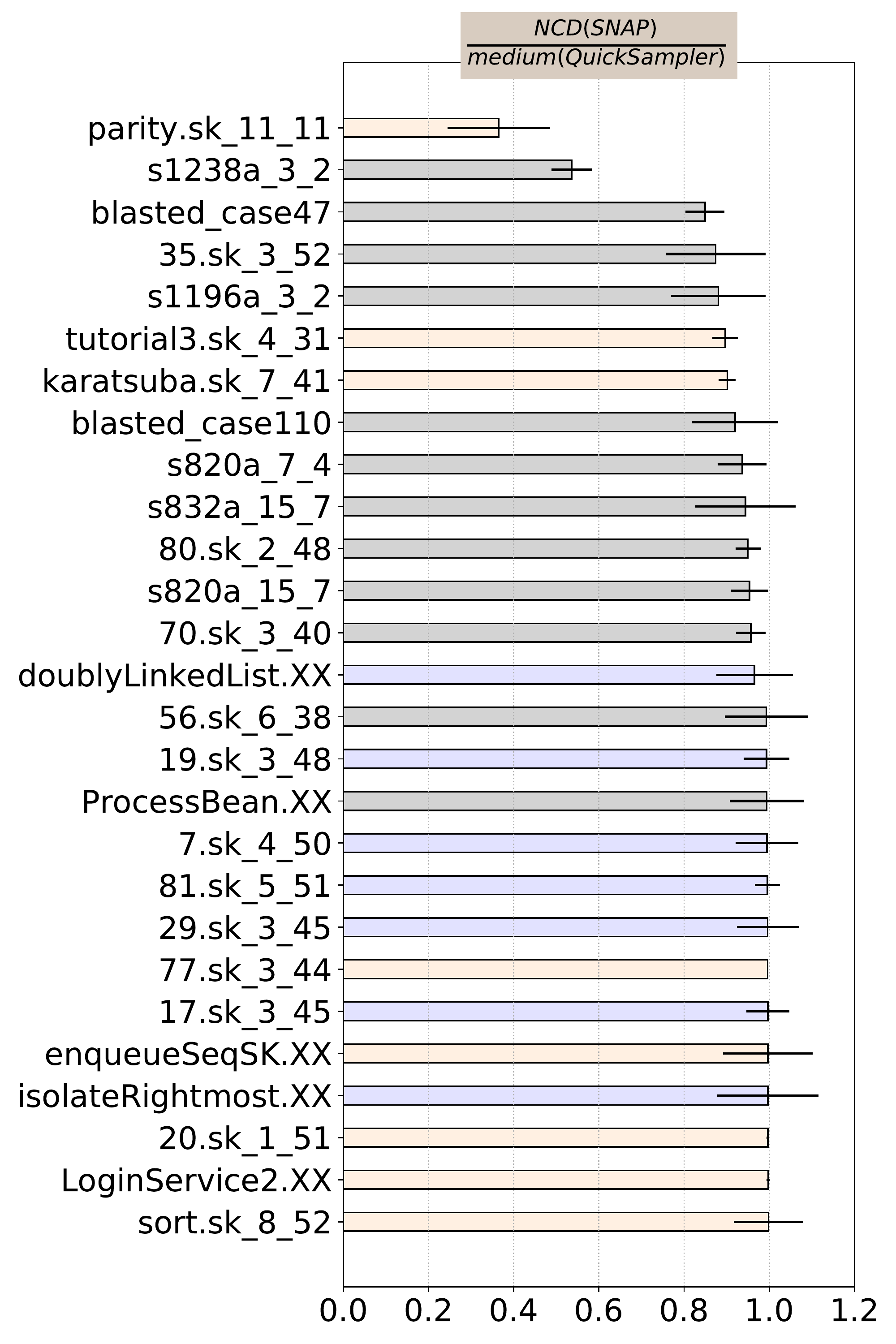}
\end{center}
\caption{RQ3 results for ``diversity'': Normalized compression distance (NCD) 
when {\em QuickSampler} and {\sc Snap}
terminated on the same case studies. 
Median results over 30 runs (and small black lines show the 75th-25th variations).
Same color scheme as \tab{benchmarks}.}\label{fig:rq21}
\end{figure}

Regarding {\em diversity},
\fig{rq21} compares the diversity
of the test suites generated by our two systems (xpressed as ratios
of the observed NCD values). Results less than one indicate that {\sc Snap}'s
test suites are less diverse than {\em QuickSampler}. In the median case, the ratio is one; i.e. in terms of central tendency, there is no difference between the two algorithms.

We have analyzed the \fig{rq21} results with a    bootstrap test at 95\% confidence (to test for statistically significant results), and a Cohen's effect size test (to rule out trivially small differences). Based on those tests, we say that in $\frac{25}{27}=93\%$ 
of these results, there is no  significant difference (of non-trivial size)
between the two algorithms.  

That said, in 
  there two cases with a   statistical significant difference
  that are markedly less than {\sc Snap}
   (see s1238a\_3\_2 and parity.sk\_11\_11) (\fig{rq21}). 
In terms of scoring different algorithms,   it could be argued that these examples might mean that {\em QuickSampler}
is the preferred algorithm but only  (a)~if  numerous invalid tests are not an issue;
(b)~if testing resources are fast and cheap (so saving time and money on cloud-compute test facilities is not worthwhile);  
and (c)~if developer time is cheap (so the time required to specify expected test output, or  processing large numbers of failed tests, is not an issue).

Hence we recommend {\sc Snap} since,
\begin{result}{3}
\BLUE
Usually,    {\sc Snap}'s tests are far more credible and 
very nearly as diverse
as those from {\em QuickSampler}.
\BLACK
\end{result}
% In summary,   we say that the {\sc Snap}
% results suggest that there is much benefit
% in assuming software shares much common structure. 
% Specifically, algorithms that make that assumption
% can ran very fast while at the same time
% generating 
% very small and very simple solutions.

\section{Threats to Validity}\label{sect:threats}

\subsection{Baseline Bias}

One threat to the validity of this work is the \textit{baseline bias}. Indeed, there are many other sampling techniques, or solvers,  that {\sc Snap} might be compared to.
However, our goal here was to compare {\sc Snap}
to a recent state-of-the-art result from {\it ICSE'18}.
In further work, we will compare {\sc Snap} to other methods.
\subsection{Internal Bias}
A second threat to validity is   {\it internal bias}
that raises from the stochastic nature of sampling techniques.
{\sc {\sc Snap}} requires many random operations.
To mitigate the threats,  we repeated the experiments for 30 times.
\BLUE

\subsection{Hyperparameter Bias}\label{sect:hpo}
Another threat is   {\it hyperparameter bias}.
The hyperparameter is the set of configurations for the algorithm.
The hyperparameter used in these experiments were shown in 
\tion{choice}.
 Learning how to automatically adjust these settings would be
 a useful direction for future work.
 
How long would it take to learn better parameters?
As shown in \fig{rq_speedup}, it can take $10^5$ (approx) seconds to complete 
one run of our test generation systems.  Standard advice for hyperparameter optimization with (say) a genetic algorithm is to mutate a population of 100 candidates over 100 generations~\cite{goldberg1988genetic}. Allowing for 20 repeats (for statistical validity), then the runtimes for  hyperparameter
optimization experiments could require:
\[10^5 * 100 * 100 * 20 / 3600 / 168 / 52   \approx 650\; \mathit{years}\]
This is clearly an upper bound. If we applied experimental hyperparameter optimizers that tried less than 50 configurations
(selected 
via   Bayesian parameter optimization~\cite{golovin2017google,nair2018finding}), then that runtimes could be
three years of CPU:
\[
10^5 * 50 * 20 / 3600 / 168 / 52   \approx 3\;\mathit{years}\]
Yet another method, that might be more promising, is incremental transfer
learning where optimizers transfer lessons learned between  hyperparameter optimizations running in parallel~\cite{9050841}. In this approach, we might not need
to wait $10^5$ seconds before we can find better parameters.

In summary, it would be an exciting and challenging task to perform hyperparameter
optimization in this domain. 

\subsection{Construct Validity}\label{sect:valid1}
There are cases where the above test scheme
would be incomplete. All the above assumes that the 
the constraints of the program can be 
expressed in terms of the literals seen within the conditionals that define each branch of a program. 
This may not always be true.  For example, consider constraints between fields of buried deep within a nested
data structure being passed around the program.
To address constraints of that type, we would need access to (e.g.) invariants that many be defined within those structs, but   which are invisible to the tests in the path conditionals.
Strongly typed languages like Haskell or OCaml which can reason about nested types might be of some assistance here. This would be a promising area for future work.
\subsection{External Validity}\label{sect:valid2}
Apart for issues of nested type constraints,
this section lists two other  areas that would require 
an extension to the current {\sc Snap} framework.
Specifically,  {\sc Snap} is  not designed for testing
{\em non-deterministic} or  {\em nonfunctional} requirements. 

{\em Functional requirements}  define  systems functions; e.g. "update credit card record".
On the other hand,
a {\em non-functional requirements} specify how the system should do it. For example,  nonfunctional requirements related to software ``ilities'' such as usability, maintainability, scalability, etc.  
When designing tests for nonfunctional   requirements, it may be required  to access variables that are not defined in the conditionals that define program
branches;  (e.g. is the user happy with the interaction?). {\sc Snap} does not do that since
it draws its tests only from the variables in the branch tests.

As to  testing non-deterministic systems,
a {\em deterministic function} is one where the output is fully determined by their inputs; i.e. if the function is called $N$ times with the same inputs then in a deterministic environment, we would expect the same output.
On the other hand,
a {\em non-deterministic function} 
is one where identical  inputs can lead to different  outputs. 
When designing  tests for non-deterministic systems,
it would be useful to make multiple tests fall down each program branch since that better samples the space of possible non-deterministic behaviours within that branch.
{\sc Snap} may not be the best tool for non-deterministic systems since, often,
it  only produces one test for each of the branches 
it visits.

In future work, it would be insightful to consider how {\sc Snap} might be extended
to handle testing for non-functional and/or non-deterministic systems.
\BLACK

\subsection{Algorithm Bias}\label{sect:abias}
One threat to the validity of the above results is that we used the
termination criteria based on NCD, which is different from some prior work.
So is  NCD
a     fair diversity comparison metric?.

\begin{figure}[!t]
\centering
\includegraphics[width=0.45\textwidth]{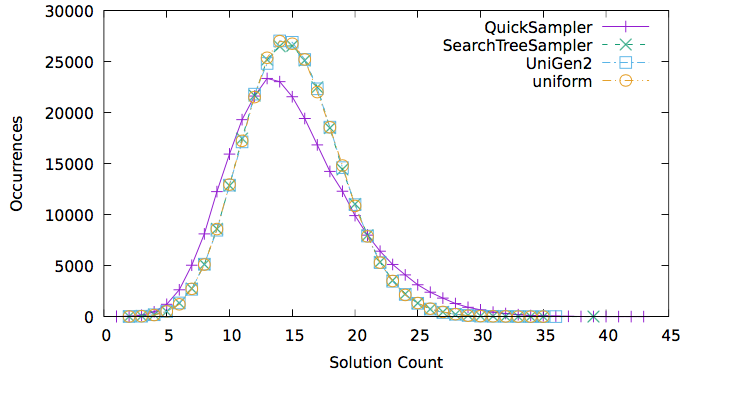}
\caption{The authors of the {\em QuickSampler} paper~\cite{dutra2018efficient},
say this figure shows that  different test generation algorithms {\em QuickSampler}\cite{dutra2018efficient}, SearchTreeSampler\cite{ermon2012uniform}, UniGen2\cite{chakraborty2015parallel}
and one uniformed random generator 
  achieve similar solution diversity
(assuming  unlimited CPU). 
%This comparison method is \textit{not} applicable to our experiment, since i) the experiment in this paper only collected unique samples ii) number of solutions from different algorithms were not in the same scale. See text for a customed 
%uniform comparison method.
}
\label{fig:old_uniform}
\end{figure}

To explore this, we need some way to compare the results
of text case generation algorithms, given the same CPU allocation, 
i.e. getting rid of NCD.
\fig{old_uniform}  shows one way to  make that comparison
  (  this figure comes from  the {\em QuickSampler} paper):
  \bi 
  \item
  Three different test case generation algorithms~\cite{dutra2018efficient,ermon2012uniform,chakraborty2015parallel} (and a uniform random generator) are executed.
  \item  Each algorithm was given 10 hours of CPU.
  \item \fig{old_uniform}  counts the number of 
repeated solutions 
within each run.
That figure shows that (e.g.)
 25,000 solutions  are found   15 times (approximately) within all four methods.
\ei
The authors of the {\em QuickSampler} paper used \fig{old_uniform}  to argue
that, assuming  a large CPU allocation,
then at the end of the run,
all these algorithms achieve similar solution diversity.
Aside: just to defend {\em QuickSampler} here--  merely because the same solutions are found in \fig{old_uniform}   by different methods does not mean that there is no benefit to {\em QuickSampler}.
As discussed in~\cite{dutra2018efficient},
{\em QuickSampler} wins over the other algorithms of \fig{old_uniform} since (a)~it scales to larger problems and (b)~it produces test suites with more valid tests, faster than previous methods.
%How can we check if {\sc Snap} achieves the
%same diversity as 
% {\em  QuickSampler}?
% Please note that those metrics and illustrations like \fig{old_uniform} was indeed an
% obvious way to compare the test suite \textbf{uniformity} among various algorithms. The  aim of {\sc Snap}, however, 
% as is addressed in \tion{why}, is to reduce the test suite size. One issue with test suite reduction is that important branches of the program might be missed.
% Therefore, to assess {\sc Snap} this evaluates the  \textbf{diversity} of test suites yielded from different generators.
To illustrate {\sc Snap}'s diversity in a similar manner to  \fig{old_uniform}, we have the following observations:
\BLACK

\bi
\item Recalling the \tab{rq4} results, {\sc Snap} produces far fewer test cases that these other algorithms. Hence, we cannot use the $y$-axis of \fig{old_uniform} to compare
our method to previous methods.
\item We need another non-NCD measure of diversity that does not
favor either {\em QuickSampler}
or {\sc Snap}. For that purpose,
 we used 
 {\it Shannon Entropy}~\cite{shannon1948mathematical}, i.e.
\[H(p) = -p\log_2^p - (1-p)\log_2^{(1-p)}\] where $p$ is the probability of one(1)s in the solution.
\ei

\begin{figure} 
\begin{center}    
\includegraphics[width=.35\textwidth]{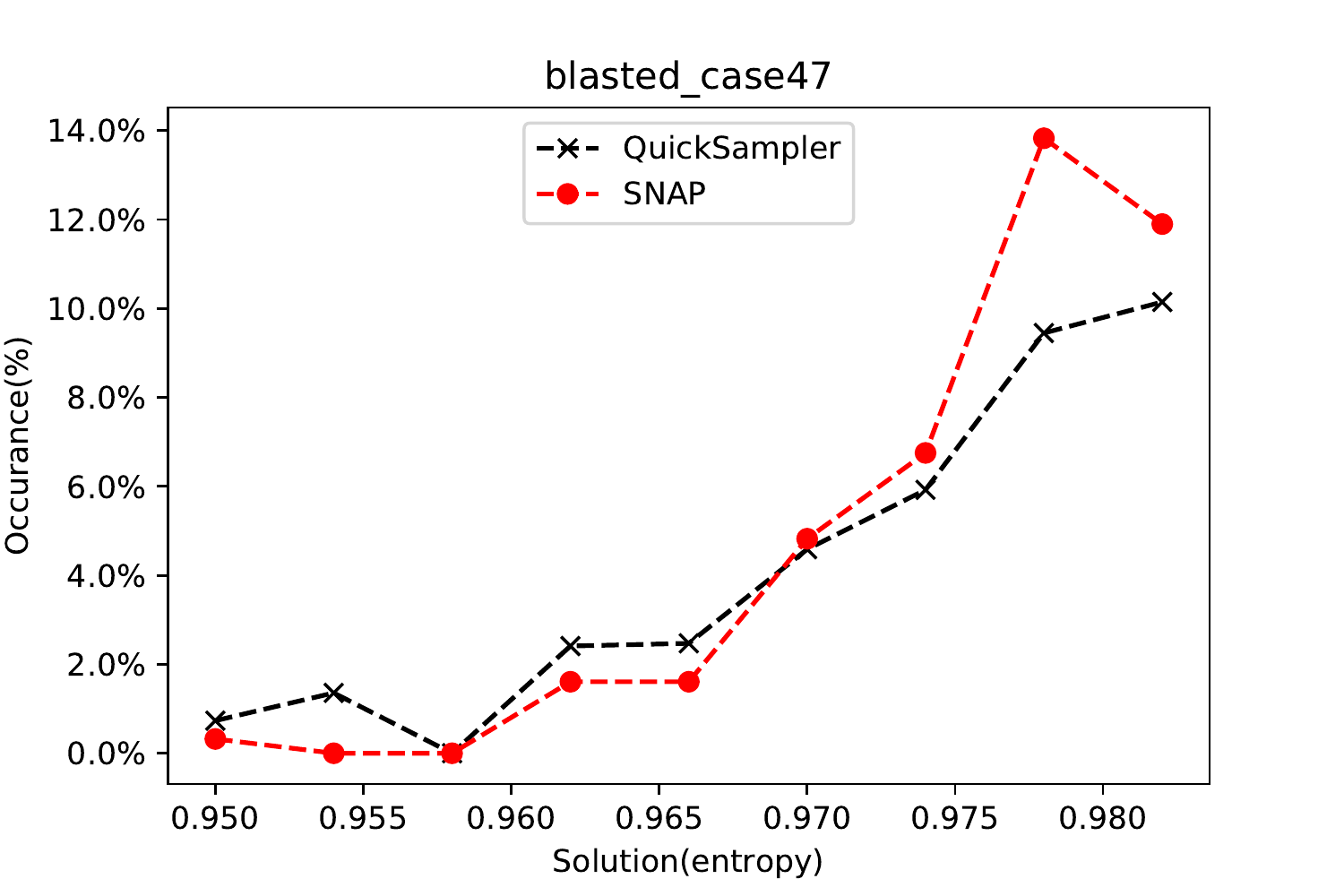}
\includegraphics[width=.35\textwidth]{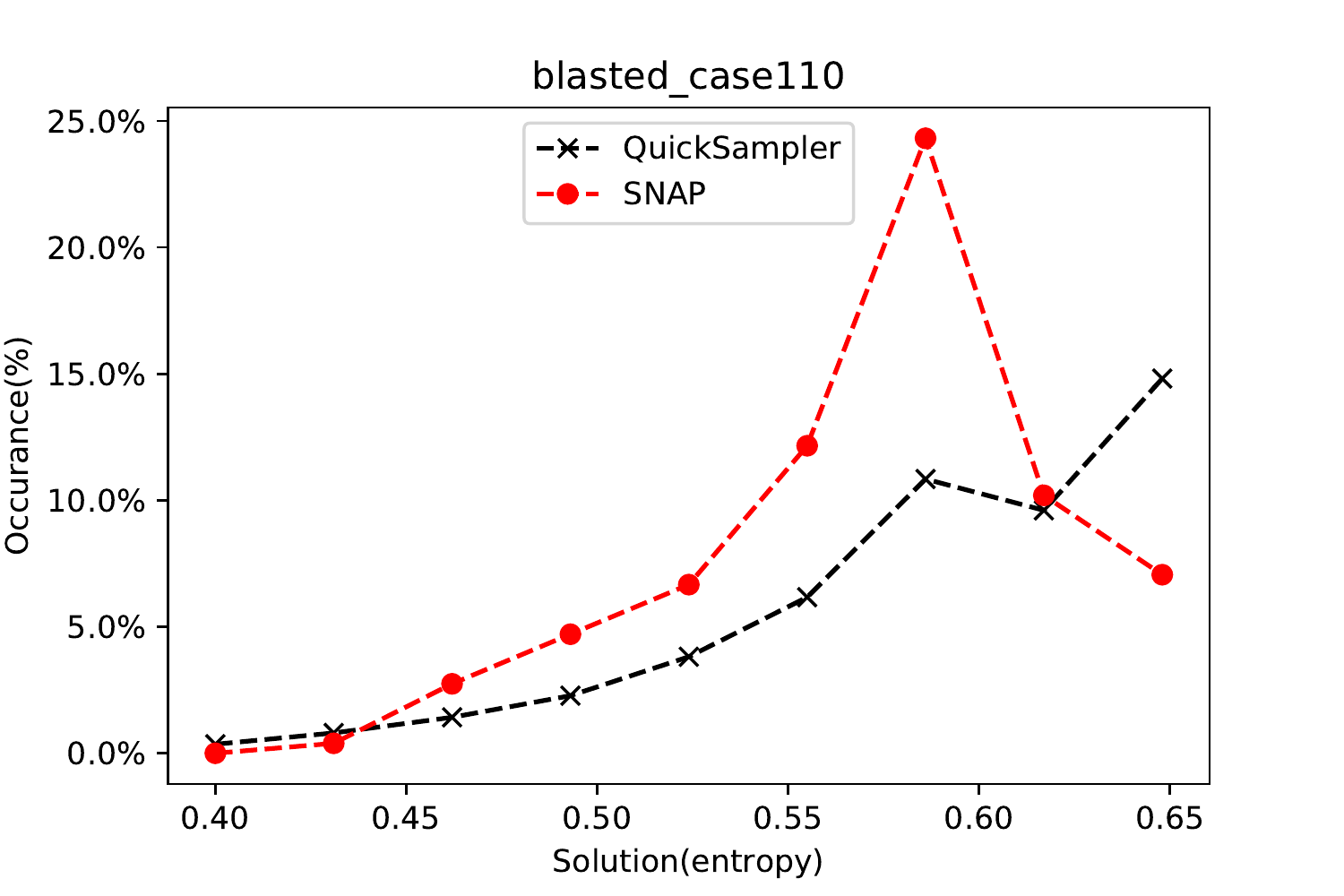}   
\includegraphics[width=.35\textwidth]{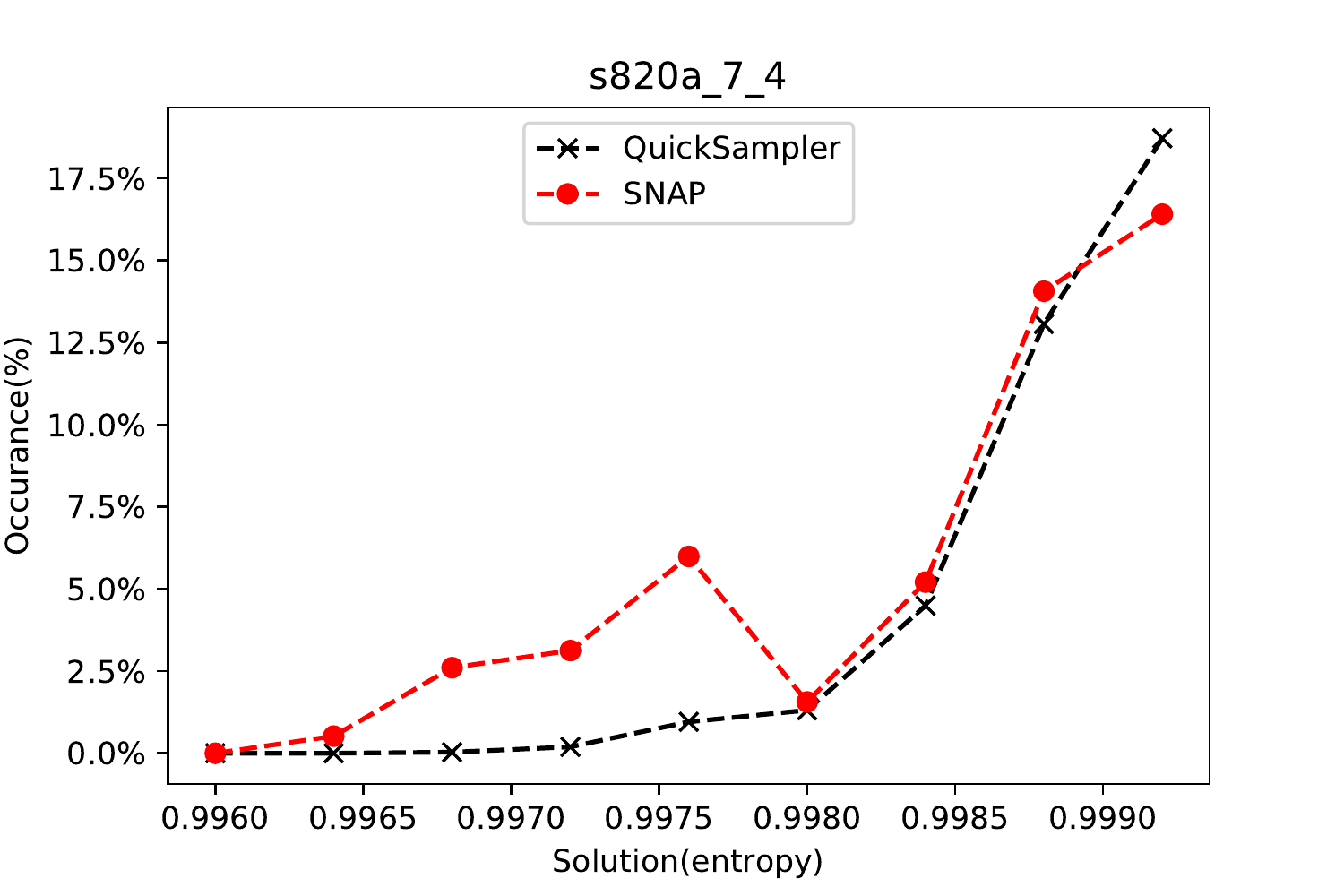}   
\includegraphics[width=.35\textwidth]{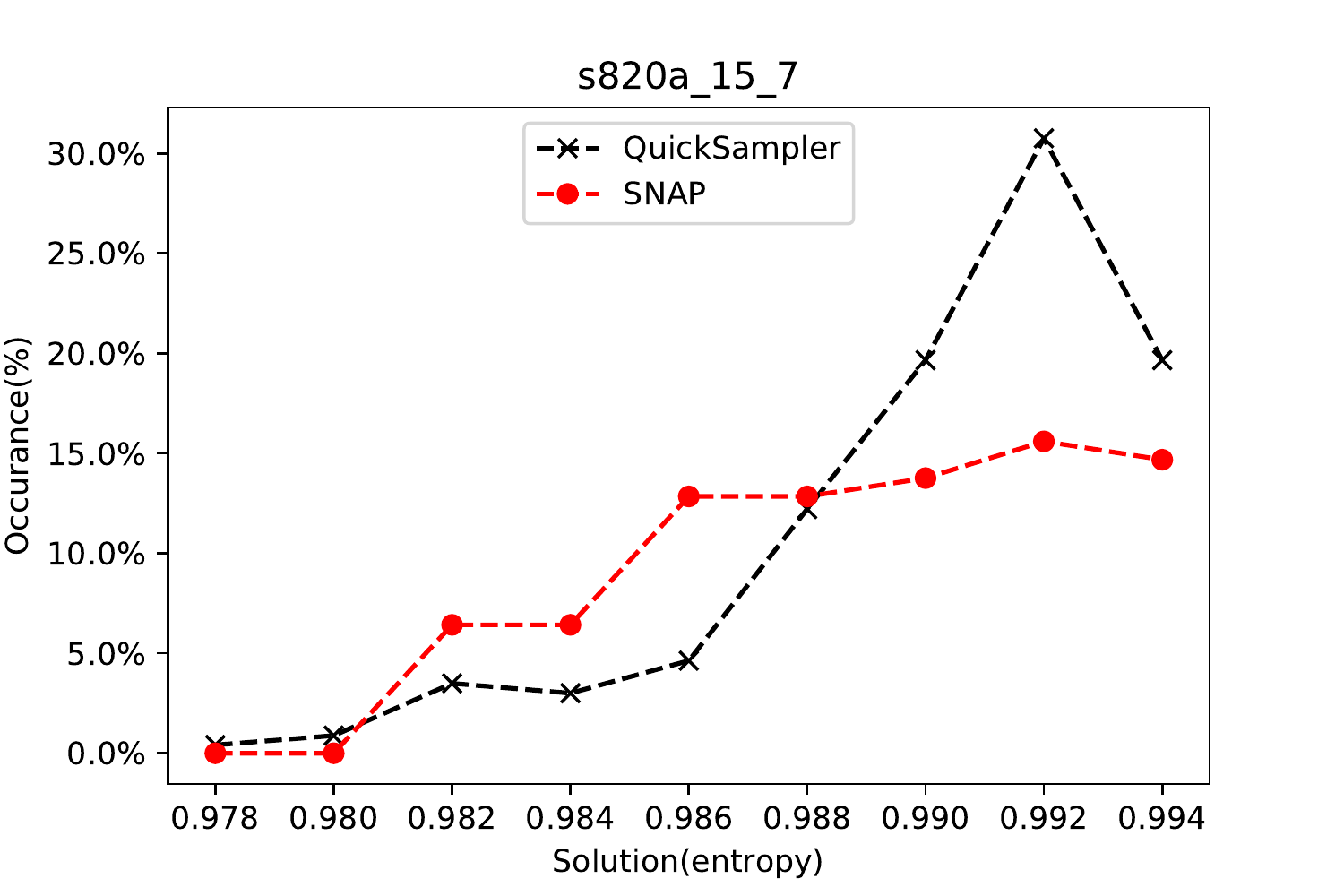} 
\includegraphics[width=.35\textwidth]{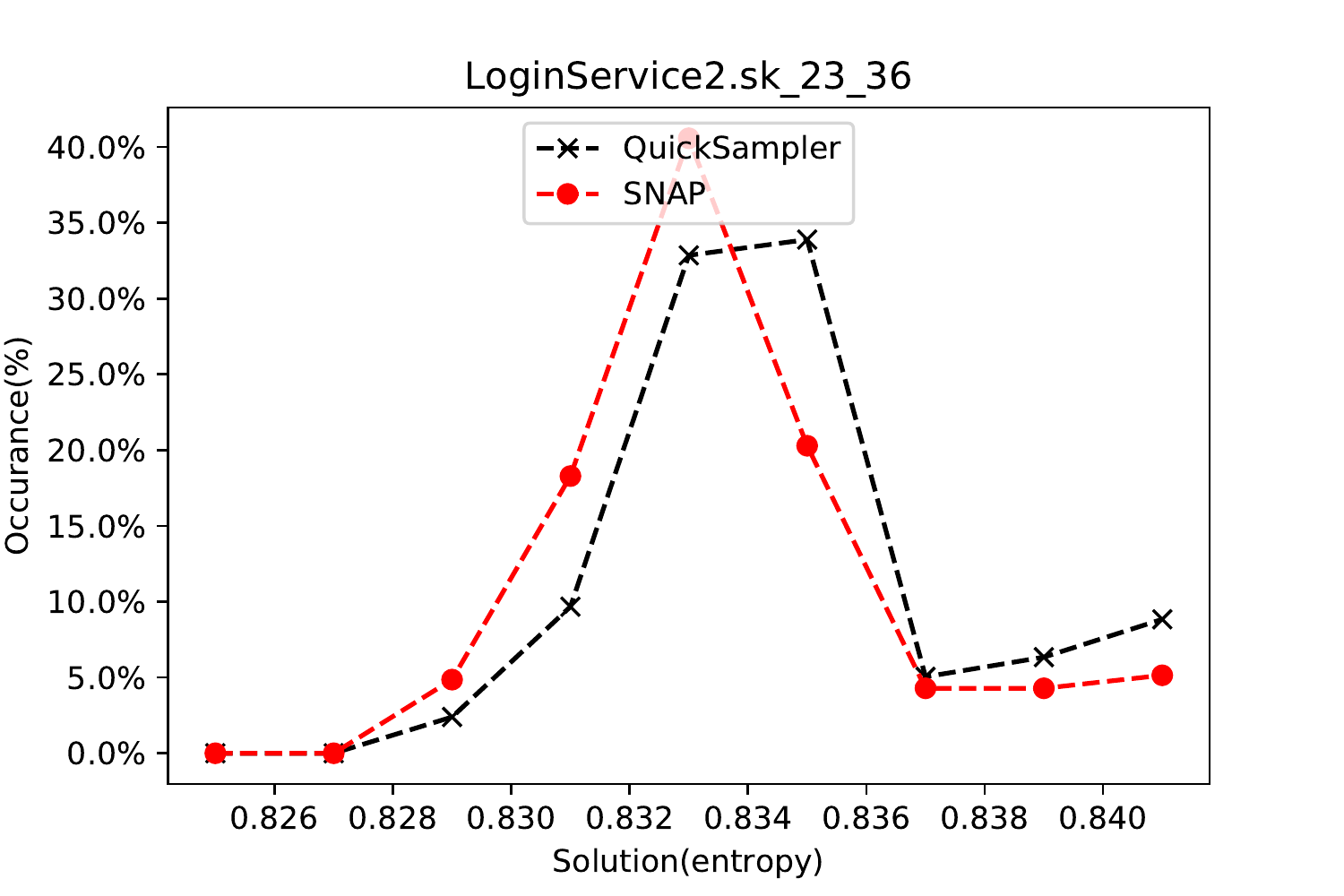} 
\end{center}
\caption{Distributions  of  the  diversity
(assessed using Shannon entropy). Both {\em QuickSampler} and {\sc Snap} were terminated at the same time -- { Blasted} cases=1min, { s820a} cases = 5mins, { LoginService2}=10mins.
$Y$-axis shows   valid solutions {\it(those with same Shannon Entropy were clustered together)}  $Y$-axis shows   occurrences among the result set.
}\label{fig:uniform_res}
\end{figure}

% \ei

% A fourth threat  is the  {\it measurement bias} used in this work. To
% to determine the diversity of a test suite, {\sc Snap} used the   normalized compression distance (NCD) described in 
% \eq{ncd} (recall that we preferred NCD to the diversity
% measure used in {\em QuickSampler} since the
% former was far faster to compute).

% Given this different measure, is the above a fair   ``apples to apples'' comparison for the two algorithms? 

In combination, our comparison proceeds as follows: 
\bi
\item
{\em QuickSampler} and {\sc Snap}
were run on each  case study,
terminating after the same  number of minutes.
\item
Since our goal was to see ``what is lost by {\sc Snap}'',
we terminated in times similar to the termination
times seen in the RQ1 study. Specifically,
those 
termination times were assigned to the case studies,
basing on their number of clauses, from the set \{1,5,10\} minutes.  
\item
As shown above in the RQ2 study,  the number of solutions generated by {\sc Snap} and {\em QuickSampler} are not in the same scale.
Accordingly, we only recorded the unique solutions found in this study.
\item
At termination, 
we collected all unique valid test cases when the execution terminated at the given time and compared the
diversities among them.
\ei
As seen in \fig{uniform_res},  we  output the distribution of entropies, expressed as
the {\em percentage of tests} that have that entropy (and not as \fig{old_uniform}'s
{\em absolute number of test }   with that entropy).

% Having applied the above procedure to dozens of our case
% studies, we can report that we did not find evidence for measurement bias due to our use of NCD.
% To document that effect, we first note that
% reviewing  these results is a somewhat extensive
% exercise. Therefore, here, we show results
% from five case studies, selected at random.
% Note that the patterns reported below
% appeared in all the other case studies we analyzed.

% \fig{old_uniform} shows the patterns of unique solutions found
% in  { blasted\_case47, blasted\_case10, s820a\_7\_4, s820a\_15\_7} and { LoginService2.sk\_23\_36}.
% Note that they are very similar.

% within hours of repeating executions of various algorithms.

\fig{uniform_res} shows the distributions of the diversity of {\sc Snap} and {\em QuickSampler} results
seen in  { blasted\_case47, blasted\_case10, s820a\_7\_4, s820a\_15\_7} and { LoginService2.sk\_23\_36}. These data sets were selected for presentation here since, in the {\em QuickSampler} paper, they were singled out for special
analysis (according to  that paper, these algorithm
yield  a large and  countable range of diverse results).
In that figure, we see that 
\bi
\item
Among all these test cases,  {\sc Snap} and {\em QuickSampler}
yielded solutions within same entropy range.
\item
In fact, 
usually we see 
{\em a very  narrow range of entropy  on the x-axis:} In 4/5 cases, the range was less than 3\%.
This   means that  these case studies yield solutions with  similar   entropy.
%Please note that
%we do not have preference on any kind of solutions. For example, we can not say solutions with higher entropy dominates
%solutions with lower entropy, {\it i.e.} the {\tt (x1, x2, x3, x4) = (true, true, true, false)} has the same value as {\tt 
%(x1, x2, x3, x4) = (true, false, true, false)}, even though the later one has higher entropy. What matters is whether or not the results can cover 
%enough range of solution entropy.
\ei
In summary, from  \fig{uniform_res}, we say that  if   solutions were  generated
 by {\em QuickSampler} with a particular entropy, 
then
it is likely that the {\sc Snap} was generating that kind of solutions as well. Hence, we do not see a threat to validity introduced by how {\sc Snap}
selects its termination criteria.

\BLUE
\subsection{Evaluation Bias}\label{sect:eval}

This paper has evaluated the {\sc Snap} test case generator using
the five goals described in the introduction:
i.e. {\em runtime},   {\em scalability},   {\em redundancy}, {\em credibility} and {\em minimality}.
But as the following 
examples show, these are  not the only
criteria for assessing test suites.  For future work it could be useful and insightful to apply
 other evaluation criteria.

Firstly, 
Yu {\it et al.}~\cite{yu2019terminator} discuss the information needs for {\em test case prioritization}. They argue that in modern complex cloud-based test environment, it can be   advantageous  not to run all tests all the time. Rather, there are engineering benefits to first running the tests that are most likely  to fail. His results show that different kinds of systems need different kinds of prioritization schemes, but not all projects collect the kinds of data needed for different prioritization  schemes. Hence it is an open issue if tools like {\sc Snap} and {\em QuickSampler} can contribute to test case prioritization.

Secondly,
once tests are run, then faults have to be {\em localized and fixed}.
Spectrum-based Reasoning (SR) is a research hot-spot on this. Given a system of $M$ components, a test suite 
$T$ as well as the obtained
errors after executing $T$ on the system, 
SR approaches utilize similarity-like
coefficient to find a correlation between component and the errors location.
% The DDU  {\it(Density-Diversity-Uniqueness)}, introduced by 
Perez {\it et al.}~\cite{Perez17} warn that though high-coverage test suites can detect errors in the system, it is not guaranteed that
inspecting tests will yield a straightforward explanation, i.e. root cause, for the error.
% Therefore, DDU estimates the density, diversity and the uniqueness and synthesizes them as a whole.
% Benchmarks for tools like {\it QuickSampler} or {\sc Snap} 
% were extracted and transformed from real-world hardware/software projects. All results were not being executed in original systems yet.
It will be  of insightful   to test how effective are  {\it QuickSampler} or {\sc Snap} in localizing faults in real-world executions.

 Thirdly, Ostrand {\it et al.}~\cite{Ostrand04} argues that the value
 of quality assurance methods is that they {\em focus the analysis} on what  parts of the code base
  deserve  most attention.   By this criteria, we should assess test suites by how well they find the {\em most} bugs in the {\em  fewest} lines of code.
 
Fourthly, a common way to assess test suite generators is via the
 {\em uniformity} of the generated tests~\cite{deng2009self}.
 Theorem provers report their solutions in some implementation-specific order.
 Hence, it is possible that after running a theorem prover for some finite time, then the solutions found in that  time may only come from a small ``corner''
 of the  space of  possible solution~\cite{chakraborty2014balancing}.
 When  test for {\em uniformity} for a  theorem prover sampling   the space of $N$ possible tests, then 
 the frequency of occurrence  of some test $T_i$ 
 should be approximately 
 $1/N$. 
 
We  argue that issues of uniformity are less important than branch coverage (which is measured
above as {\em diversity}, see \tion{abias}).  
To make that argument, we draw
 a   parallel from the field of data mining. Consider a rule learner that is building a rule from the set of all possible literals in a  data sets.
  In theory,
 this space of literals is very large
 (all attributes combined with all their ranges combined any any number of logical operators and combined to any length of rule). 
 Nevertheless, a repeated result is that 
 such learners 
 can terminate very quickly~\cite{hand2014data} since,
 rather that searching 
 all literals, these
 learners need only explore  the small set of literals 
 commonly  seen
 in the data. 

We draw this parallel since 
the success of {\sc Snap} is consistent with the conjecture that the programs we  explore are using just a small subset of the space of all  settings.
 In that situation,  uniformity is less of an issue than diversity since the latter reports how well the tests match the ``shape''  of the data. 

We note that other researchers endorse our position here that effective testing need only explore a small portion of the total state space.   Miryung Kim and colleagues~\cite{kim19zz} were  testing scripts that processed up to $10^{10}$
rows of data. In theory,  the
test suite required here is very    large indeed (the cross-produce of all the possible values in
$10^{10}$ rows). However, a static analysis   showed that those scripts could be approximated by less than 3 dozens pathways. Hence in that application, less than three dozen tests were enough to test those scripts. 
Note the parallels of the Kim et al. results to  the {\sc Snap} work and the data mining example
offered above: 
\bi
\item
Kim et al. did not cover all possible data combinations. 
\item
Rather, they constrained their tests to just cover
the standard ``shape'' of the code they were testing.
\ei
Accordingly,  when seeking  the smallest number of tests that cover the branches, it may be a secondary concern whether or not those test cases ``bunch up'' and do not cover the cross product of all possible solutions.

\BLACK
\section{Related Work}\label{sect:backdoors}
  
In essence, the algorithms of this paper are {\em samplers} that explore some subset of seemingly large and complex problems.
Sampling is not only useful for  finding test suites in theorem proving.
It also has applications for other 
SE problems such as requirement engineering,
resource planning optimization, etc~\cite{chen2018sampling,chen2018beyond,menzies2005xomo,chen2018riot}.
A repeated problem
with all these  applications was the time required
to initialize the reasoning. In that initialization step, some large number of
samples had to be collected. In practice, that step took a significant
percentage of the total runtime of those systems.
We conjecture that {\sc Snap} can solve that initialization problem.
Using the techniques of this paper, it might be time now to repeat all the above work. This time, however, instead of wasting much time on a tedious generation process, we could use something like {\sc Snap} to quick start the reasoning.

As to other related work,
 like {\sc Snap}, the DODGE system of Agrawal {\it et al.}~\cite{Agrawal19} made an assumption that 
given a set of solutions to some SE problem, there is  much redundancy and repetition within those different solutions.
  A tool for software analytics, DODGE   needed just a few dozen evaluations to explore billions of    configuration options 
 for (a)~choice of learner, for (b)~choice of pre-processor, 
 and for (c)~control parameters for the learner and pre-processor.
 DODGE executed by:
 \be
 \item
 Assign   random weights to   configuration options.
 \item
 Randomly pick  
  options, favoring   those with most weight;
  \item
  Configuring and executing data pre-processors and learners using those options;
  \item
  Dividing output scores into regions of size $\epsilon=0.2$;
 \item
 When some new configuration has scores with  $\epsilon$ of  
  prior configurations then...
  \item
  ...reduce the weight of those configuration options; 
  \item Go     Step2
  \ee
  Note that after  Step5, then the choices made in subsequent Step1s will avoid options that arrive within $\epsilon$ of 
 other observed scores. Experiments with DODGE found that best learner performance plateau after just a few dozen repeats of Steps12345.
 To explain this result,
Argrawal {\it et al.}~\cite{Agrawal19} note that for a range of software analytics tasks,  the outputs of a learner divide into  only a handful of equivalent regions.
For example, when an software analytics task
  is repeated 10 times, each time with 90\% of the
data, then the  observed performance scores   (e.g. recall, false alarm)
can vary by 5 percent, or more. Assuming normality, 
then  scores less than $\epsilon=1.96*2*0.05=0.196$ are statistically indistinguishable. Hence, for learners    evaluated on (say) $N=2$  scores, 
  those  scores   effectively divide into just \mbox{ $C=\left(\frac{1}{\epsilon = 0.196}\right)^{N=2}=26$} different regions.  Hence, it is hardly surprising that a few dozen repeats of Step1,2,3,4,5 were enough to explore a seemingly very large space of options.

It has not escaped our notice that some analogy of the  DODGE result could explain the curious success of the QuickSampler heuristic.
Consider: 
one way to summarize  \eq{one}. 
is  that the space around existing valid test cases
contains many other valid test cases-- which is an analogous
idea to   Argrawal's
$\epsilon$ regions.
That said, we would be hard pressed to defend that analogy. 
 Argrawal's
$\epsilon$ regions are a statistical concept based on continuous variables while  \eq{one} is defined over discrete values.

Also, there are many other ways in which DODGE is
fundamentally different to {\sc Snap}.
DODGE was a support tool for {\em inductive} data mining applications while {\sc Snap} is most accurately described as a support tool for a {\em deductive} system (Z3). 
Further, DODGE assumes very little structure in its inputs (just tables of data with no more than a few dozen attributes) while {\sc Snap}'s inputs are far larger and far more structured (recall from Table~\ref{tab:benchmarks} that {\sc Snap} processes CNF formula with up to hundreds of thousands of variables).
Lastly,
recalling Step6 (listed above), DODGE incrementally re-weights the space from which new options are generated. {\sc Snap}, on the other hand, treats the option generator 
as a black box algorithm since it does not reach inside Z3 to
alter the order in which it generates solutions.

\section{Conclusion}\label{sect:conclusion}

Exploring propositional formula is a core computational process with many areas
of application. Here, we explore the use of such formula for test suite generation.
SAT solvers are a promising technology for finding settings that satisfy
propositional formula. The   
  current generation of SAT solvers is challenged by the size of the formula seen in the recent SE testing literature.

Using the criteria listed in the introduction 
({\em runtime},   {\em scalability},   {\em redundancy}, {\em credibility} and {\em minimality}),  we recommend the following ``{\sc {\sc Snap}} tactic''  to tame the computational complexity of SAT solving 
for test suite generation:
\begin{quote}
{\em Sample around the average values seen in  a few randomly selected valid tests.}
\end{quote}
When this tactic was  applied to 27 real-world test case studies, test suite generation can ran 10 to 3000 times faster (median to max) than a prior report. While that prior work found tests that were 70\% valid, our  {\sc Snap} tool generated 100\% valid tests. 

Another important result was the size of the test set generated in this manner.
There is an economic imperative to
run fewer tests when companies have to pay money to run each test, and when developers have to spend time
studying the failed test.  In that context, it is interesting  to note that  
Snap's tests are 10 to 750 times smaller (median to max) than those from prior work.

We conjecture that:
\bi
\item
{\sc Snap}'s success is due to  widespread repeated structures in software.
Without such repeated structures, we are at a loss to explain our results. Algorithms that exploit such repeated  structures,
there are many kinds of SE analysis that (potentially) could be clarified and simplified using 
   {\sc Snap}. 
 \item
Given the presence of such repeated strucutures, the  {\sc Snap} tactic might be useful for many other SE tasks.
\ei
\section*{Acknowledgements}
This work was partially funded by an NSF
  award \#1703487. 
  %Any opinions, findings, and conclusions or recommendations expressed in this
%material are those of the authors and do not necessarily reflect the
%views of NSF.

\balance
\bibliographystyle{IEEEtran}
\input{main.bbl}

\begin{IEEEbiography}[{\includegraphics[width=1.05in,clip,keepaspectratio]{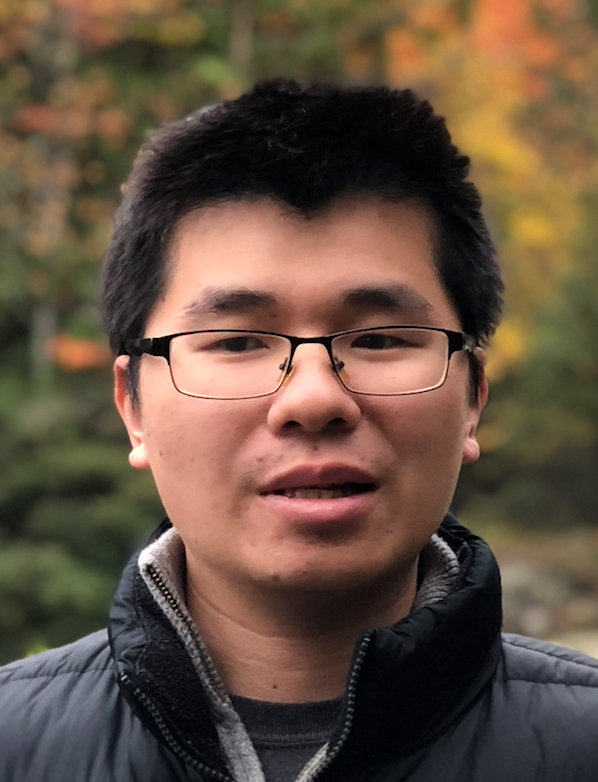}}]{Jianfeng Chen } holds a Computer Science Ph.D. from North Carolina State University. His research     mainly focuses on utilizing machine learning techniques to optimize software engineering/system infrastructure performance. He
 works now as a research scientist
at Facebook, New York.
For more information,  please visit
\url{https://jianfeng.us/}.
\end{IEEEbiography}

\begin{IEEEbiography}[{\includegraphics[width=1.05in,clip,keepaspectratio]{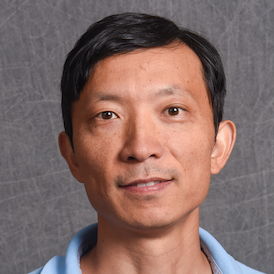}}]{Xipeng Shen} is a Professor in the Computer Science Department at North Carolina State University in USA. He is a receipt of the DOE Early Career Award, NSF CAREER Award, Google Faculty Research Award, and IBM CAS Faculty Fellow Award. He is an ACM Distinguished Member, ACM Distinguished Speaker, and a senior member of IEEE. He was honored with University Faculty Scholars Award. His primary research interest lies in the fields of Programming Systems and Machine Learning, emphasizing inter-disciplinary problems and approaches. 
\end{IEEEbiography}

\begin{IEEEbiography}[{\includegraphics[width=1.05in,clip,keepaspectratio]{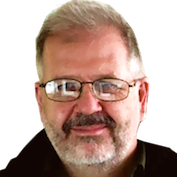}}]{Tim Menzies} (IEEE Fellow, Ph.D. UNSW, 1995)
is a Professor in computer science  at NC State University, USA,  
where he teaches software engineering,
automated software engineering,
and programming languages.
His research interests include software engineering (SE), data mining, artificial intelligence, and search-based SE, open access science. 
For more information,  please visit \url{http://menzies.us}.
\end{IEEEbiography}
\end{document}

%% file: main.bbl
% Generated by IEEEtran.bst, version: 1.14 (2015/08/26)